\documentclass[11pt]{article}
\usepackage{eurosym}
\usepackage{graphicx}
\usepackage[utf8]{inputenc}
\usepackage[T1]{fontenc}
\usepackage{indentfirst}
\usepackage[margin=0.6in,nomarginpar]{geometry}
\usepackage[final]{hyperref}
\usepackage{amsmath}
\usepackage{hyperref}
\usepackage{cite}
\usepackage{subcaption}
\usepackage{caption}
\usepackage{amssymb}
\usepackage{multirow}
\usepackage[table]{xcolor}
\usepackage{orcidlink}
\usepackage{multirow,stackengine}

\hypersetup{colorlinks=true, linkcolor=blue, citecolor=blue, urlcolor=blue}
\begin{document}

\date{\today }

\begin{center}
    {\bf Euler-Heisenberg black hole  surrounded by quintessence in the background of perfect fluid dark matter: Thermodynamics, Shadows and Quasinormal modes}
\end{center}

\vspace{0.1cm}

\begin{center}
        {\bf B. Hamil\orcidlink{0000-0002-7043-6104}}\footnote{\bf hamilbilel@gmail.com/bilel.hamil@umc.edu.dz }\\
    { Laboratoire de Physique Math\'{e}matique et Subatomique,\\
Facult\'{e} des Sciences Exactes, Universit\'{e} Constantine 1, Constantine,
Algeria}\\
    \vspace{0.15cm}
    {\bf B. C. L\"{u}tf\"{u}o\u{g}lu\orcidlink{0000-0001-6467-5005}}\footnote{\bf bekir.lutfuoglu@uhk.cz (Corresponding author)}\\
    { Department of Physics, Faculty of Science, University of Hradec Kralove,\\
Rokitanskeho 62/26, Hradec Kralove, 500 03, Czech Republic.}\\
    \vspace{0.15cm}
    
\end{center}
\date{\today} 

\vspace{0.3cm}

\begin{abstract}
Current observations show that a significant fraction of the Universe is composed of dark energy and dark matter. In this paper, we investigate the simultaneous effects of these dark sectors on the Euler-Heisenberg black hole, using the quintessence matter field and perfect fluid to model them.  In particular, we study the black hole's thermodynamics, shadows, and quasinormal modes, and discuss in detail how these properties change with relatively large or small dark sector components. 
\end{abstract}

\vspace{0.1cm}

\section{Introduction}
In 1915, Einstein introduced his revolutionary general theory of relativity, which suggests that gravity is a geometric property of spacetime \cite{Einstein}. Since then, general relativity has been considered the most accurate theory for describing the phenomenon of gravity. One set of solutions to Einstein's equations in general relativity predicts the existence of black holes. A black hole is one of the most enigmatic phenomena of the Universe, which produces an effective area from which even light cannot escape. Through the groundbreaking work of Hawking and Bekenstein on black hole temperature and entropy, it was revealed that there exists a profound connection among gravity, quantum theory, and thermodynamics \cite{Bekenstein,Hawking, Hawking1,Bardeen}.As is known, the four fundamental laws of thermodynamics have significant parallels in black hole physics. For instance, the second law of thermodynamics is similar to the area law of black hole dynamics, indicating that the surface area of a black hole cannot decrease. Additionally, the Hawking temperature, entropy, and mass of black holes satisfy the first law of black hole thermodynamics, drawing a parallel between the two domains.%

One of the most fundamental features of the current Universe is its accelerated expansion. Recent cosmological observations, such as Type Ia supernovae \cite{Riess,Ries,Rss,Knop} cosmic microwave background radiation anisotropies \cite{Miller,Hanany,Mason,Page}, and the large-scale structure of the Universe \cite{Scranton,Tegmark}, strongly support this accelerated expansion. However, the cause of this accelerated cosmic expansion remains unknown. According to Einstein's theory of gravity, this acceleration can be explained by the hypothesis that, in the later stages of the Universe, a considerable part of its total energy density is dominated by dark energy with negative pressure \cite{Peebles,Padmanabhan,Padmanabhan1}. There are two candidates for the dark energy. The first and the simplest explanation for dark energy is the cosmological constant \cite{Weinberg}  and the second is the so called quintessence \cite{Carroll},  with the state equation $p_{q}=\omega_{q} \rho_{q}$ , where $p_{q}$ is the pressure, $\rho_{q}$  is the energy density and $\omega_{q}$ in the range of $ -1<\omega_{q}<-1/3$. The special case of $\omega_{q}=-1$ for quintessence corresponds to the cosmological constant scenario.

Indirect but reliable evidence, such as the rotation curves of spiral galaxies \cite{Rubin1970, Roberts1973, Rubin1980}, the dynamical stability of disk galaxies \cite{Einasto1974, Ostriker1973}, the formation of large-scale structure in the Universe \cite{Davis1985}, gravitational lensing observations of galaxies and galaxy clusters \cite{Clowe2006, Mandel2013}, the cosmic microwave background \cite{Hinshaw2013, Ade2014}, and baryon acoustic oscillations \cite{Komatsu2011},  points out that a considerable part of the Universe is made up of dark matter, distinct from ordinary observable matter and dark energy \cite{Bertone2018, Huterer2018}. Therefore, especially in the last two decades, scientists have proposed many theories to model dark matter, which goes or not beyond the Standard Model \cite{Navarro1996, Boehm2004, Bertone2005, Feng2009, Graham2015, Schumann2019, Boyarsky2019, Qiao2021, Deur2019, Yaholom2020, Oks2020, Oks2024}. So far, cold, warm and scalar field dark matter models have been considered as suitable dark matter models \cite{Ruiz2021, Abbas2023, Chen2024}. In addition to these, a new model, {perfect fluid dark matter (PFDM) \cite{Kiselev2002, Kiselev2003, Sidd2003},} has been added to the list of suitable models because of its success in describing the asymptotically flat rotation curves of observed spiral galaxies \cite{Rahaman2010, Potapov2016}. Although in some aspects the real dark matter could not fulfill the perfect fluid condition, still it has been widely assumed as a consistent approach, and it has been used widely especially in recent years \cite{Li2012, Xu2018, Hou2018, Haroon2019, Rizwan2019, Hendi2020, Zhang2021, Atamurotov2021, Rayimbaev2021, Atamurotov2022, Ndongmo2023, Rakhimova2023, Heydarifard2023,
Qiao2023, Qi2023, Das2024, Sadeghi2024, Ma2024,  Sood2024, Biz2024}.

On the other hand, one of the major challenges in general theory of relativity is the singularities, which occur both at at the beginning of the Universe and at the center of black holes. Similarly, Maxwell's equations, which govern electromagnetism, also exhibit singularities that lead to divergence issues within Maxwell's theory. Inspired by Dirac's theory of electrons-positrons, Heisenberg and Euler proposed a novel approach to describe the electromagnetic field. Their approach incorporated one-loop corrections to quantum electrodynamics (QED) and explains the phenomenon of vacuum polarization in QED \cite{Heisenberg}. Later, Schwinger in his famous paper \cite{Schwinger}, applied a renormalization of field strength and charge to the modified Lagrange function for constant fields. Yajima and Tamaki derived the Euler-Heisenberg black hole solution by examining the one-loop effective Lagrangian density coupled with the Einstein field equation \cite{Yajima}. Kruglov consider Heisenberg-Euler-type model of nonlinear electrodynamics involves two independent parameters. Within this framework, he derived a solution for a charged black hole in the context of nonlinear electrodynamics. Additionally, Kruglov obtained corrections to Coulomb's law of electrostatics in the limit of large distances. Furthermore, he examined the energy conditions and demonstrated that the weak energy condition, dominant energy condition and  strong energy condition are satisfied for the case $B = 0$, $E \neq 0$ \cite{Kruglov}. Recently there has been extensive research into the properties of black holes within the framework of nonlinear electrodynamics. Moreover, significant attention has been given to investigating how nonlinear electrodynamics affects thermodynamic phase transitions and the shadows of black holes \cite{Ruffini,Macias,Alfredo,Breton,Marco,Zhao,Zeng,Sen,Qing, Guerrero2020, Amaro2020, Allahyari2020, Breton2021, Yu2023, Rehman2023, Magos2023, Mushtaq2024}

These theoretical speculations, together with recent astronomical observations, motivate us to study the effect of the dark sector of the Universe on Euler-Heisenberg black holes in the contexts of thermodynamics, shadows and quasinormal modes (QNM). The structure of the manuscript is as follows: In section \ref{sec:2}, we provide a brief review of the Euler-Heisenberg theory and derive the black hole solution in the presence of quintessence field within  PFDM background. In section \ref{sec:3}, we investigate the thermodynamics of the black hole. Then, in section \ref{sec:4}, we study the black hole shadows. Next, in section \ref{sec:5}, we calculate the quasinormal modes. In the subsequent section, we briefly summarize our findings to conclude the manuscript.

\section{A brief review} \label{sec:2}
In this section, we provide an overview of the Euler-Heisenberg theory combined with gravity. We first express the four-dimensional action of general relativity coupled with nonlinear electrodynamics (NLED) \cite{Plebanski, Salazar}
\begin{equation}
\mathbf{S=}\frac{1}{4\pi }\int d^{4}x\sqrt{-g}\bigg[ \frac{1}{4}R-\mathcal{L}%
( F,G)\bigg] ,
\end{equation}%
where $R$ represents the Ricci scalar curvature, and $\mathcal{L}(
F,G) $ denotes the Lagrangian density for the Euler-Heisenberg-NLED,%
\begin{equation}
\mathcal{L}\left( F,G\right) =-F+\frac{a}{2}F^{2}+\frac{7a}{8}G^{2},
\end{equation}
{ while 
\begin{eqnarray}
 F=\frac{1}{4}F_{\mu \nu }F^{\mu \nu } \qquad \text{and} \qquad  G=\frac{1}{4}F_{\mu \nu }^{\text{ \ \ }\ast }F^{\mu \nu }.
\end{eqnarray}
Here, $a$ is the Euler-Heisenberg parameter that regulates the strength of the NLED contribution and it is defined with the fine structure constant, $\alpha _{fs}$, and the electron mass, $m$,  in the natural units ($\hbar=c=1$) as follows:
\begin{equation}
a=\frac{8\alpha _{fs}}{45m^{4}}.
\end{equation}
}
The equations of motion derived from this action are more conveniently expressed using the Legendre dual description of NLED, which was introduced by Plebański in \cite{Plebanski}. This approach involves the introduction of a tensor $P_{\mu \nu}$,
\begin{equation}
P_{\mu \nu }=\left( 1-aF\right) F_{\mu \nu }-^{\ast }F_{\mu \nu }\frac{7a}{4}G,
\end{equation}%
which corresponds to the electric induction vector $%
\mathbf{D}$ and the magnetic field $\mathbf{H}$. Within the $P$ framework,
there are two independent invariants, $s$ and $t$, defined as follows:%
\begin{equation}
s=-\frac{1}{4}P_{\mu \nu }P^{\mu \nu }\text{ \ \ and \ \ }t=-\frac{1}{4}%
P_{\mu \nu }^{\text{ \ \ }\ast }P^{\mu \nu }.
\end{equation}
{ In circumstances where the electric fields are relatively weak in comparison to the critical field strength $E_{c}$, 
\begin{equation}
 \frac{\alpha _{fs}}{E_{c}}<<1  , 
\end{equation}
a perturbative expansion of the Euler-Heisenberg Lagrangian becomes a valid approach. Since $a$ is small, higher-order corrections in $a$ are negligible compared to the first-order term. It is important to note that this assumption is always valid unless the system is subjected to extreme conditions, such as the presence of very strong fields or highly charged black holes. Based on this premise, we express the Hamiltonian in the first order in $a$ as follows: }
\begin{equation}
\mathcal{H}=s-\frac{a}{2}s^{2}-\frac{7a}{8}t^{2}.
\end{equation}
In \cite{Salazar}, the equations of the motion for the coupled system are given by 
\begin{equation}
\nabla _{\mu }P^{\mu \nu }=0,\text{ \ \ \ \ }R_{\mu \nu }-\frac{1}{2}g_{\mu
\nu }R=8\pi \Bar{T}_{\mu \nu },
\end{equation}
with the energy-momentum tensor
\begin{equation}
\Bar{T}_{\mu \nu }=\frac{1}{4\pi }\left[ \left( 1-aP\right) P_{\mu }^{\sigma
}P_{\nu \sigma }+g_{\mu \nu }\left( s-\frac{3a}{2}s^{2}-\frac{7a}{2}%
t^{2}\right) \right] .\label{mt}
\end{equation}
If one sets $a=0$, Eq. (\ref{mt}) reduces to the standard energy-momentum tensor for linear Maxwell theory.%

\begin{figure}[htb!]
\begin{minipage}[t]{0.35\textwidth}
        \centering
        \subcaption{{$\omega_{q}=-0.35$} }
        \includegraphics[width=\textwidth]{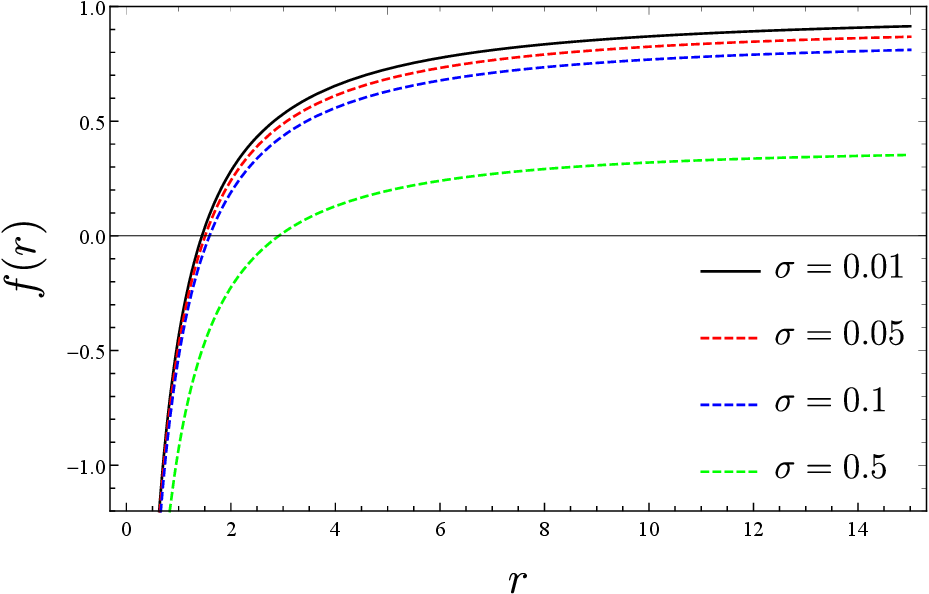}
         \label{fig:phb}
\end{minipage}%
\begin{minipage}[t]{0.35\textwidth}
        \centering
        \subcaption{{ $\omega_{q}=-0.65$}}
        \includegraphics[width=\textwidth]{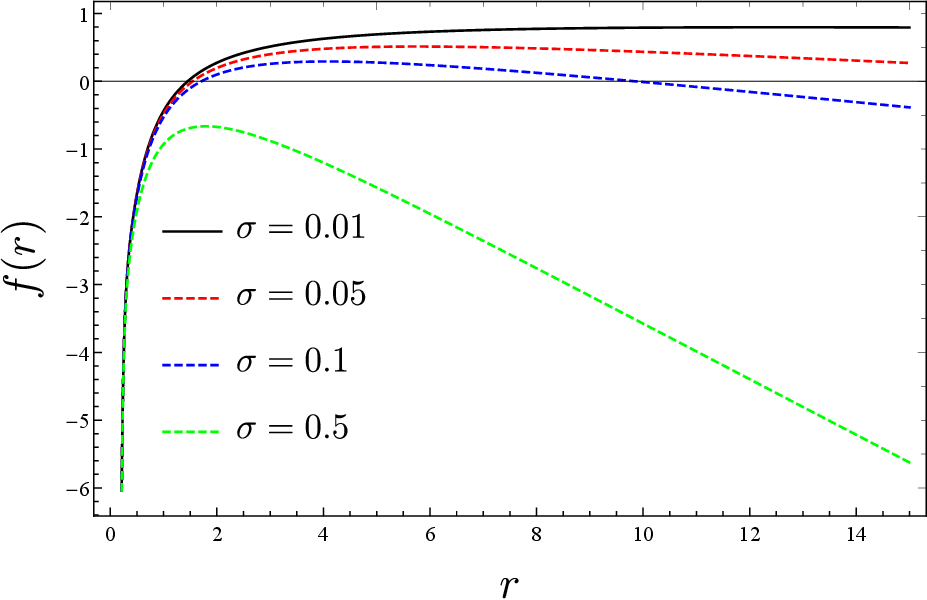}
       \label{fig:phc}
   \end{minipage}%
\begin{minipage}[t]{0.35\textwidth}
        \centering
        \subcaption{ {$\omega_{q}=-0.95$}}
        \includegraphics[width=\textwidth]{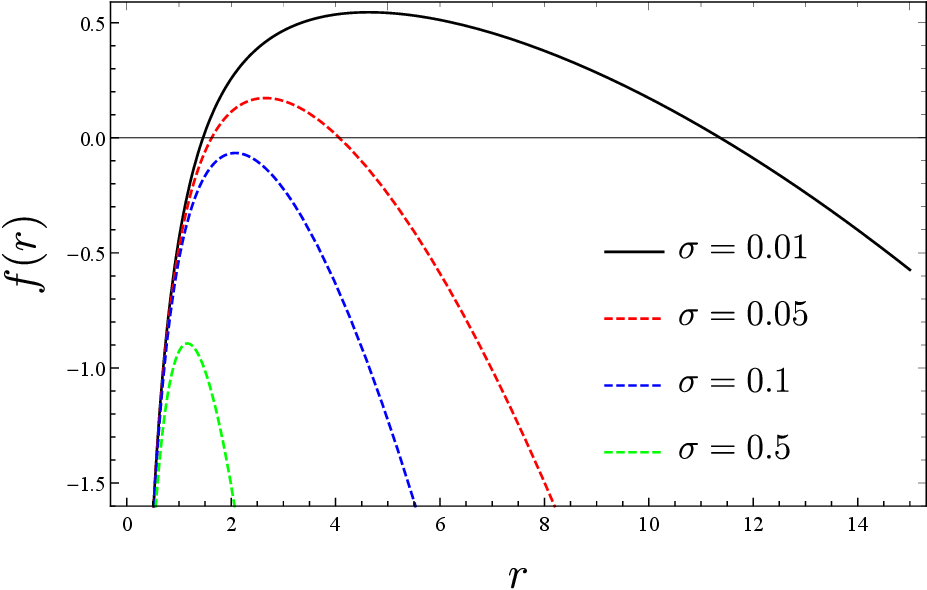}
         \label{fig:phd}
   \end{minipage}
\caption{ {{\footnotesize The impact of the quintessence field on the lapse function of the Euler-Heisenberg black hole for $M=1$, $q=0.5$, $a=0.1$ and $\alpha=0.2$.}}}
\label{lafun}
\end{figure}

Now, we handle the structure in the presence of quintessence and PFDM.  In this case, the action corresponding to the Einstein gravity coupled to Euler-Heisenberg reads \cite{Li2012}
\begin{equation}
\mathbf{S=}\frac{1}{4\pi }\int d^{4}x\sqrt{-g}\left[ \frac{1}{4}R-\left( 
\mathcal{L}\left( F,G\right) +4\pi {\mathcal{L}_{\rm DM}}-{\mathcal{L}_{\rm qui}}\right) %
\right].\label{act2}
\end{equation}
Here, the dark matter contribution to the action is represented by the term $\mathcal{L}_{\rm DM}$, which is the Lagrangian density for dark matter. The quintessence term is denoted by $\mathcal{L}_{\rm qui}$ and given in the form of \cite{Mehdi}:%
\begin{equation}
{\mathcal{L}_{\rm qui}}=-\frac{1}{2}\left( \nabla \phi \right) ^{2}-V\left( \phi
\right),
\end{equation}
with  the quintessence scalar field, $\phi$, and  the potential term associated with the quintessence field, $V\left( \phi\right) $. In this case, the equations of the motion for the coupled system reads
{
\begin{equation}
\nabla _{\mu }P^{\mu \nu }=0,\text{ \ \ \ \ }R_{\mu \nu }-\frac{1}{2}g_{\mu
\nu }R=8\pi (\Bar{T}_{\mu \nu }+4\pi T^{(\rm DM)}_{\mu\nu}-T^{(\rm qui)}_{\mu\nu}),\label{ef}
\end{equation}
}
where the energy-momentum tensor of the quintessence field and the PFDM stand with the following forms, respectively:
{
\begin{eqnarray}
T^{(\rm qui)}_{\mu\nu}&=&\nabla _{\mu }\phi \nabla _{\nu }\phi -\frac{1}{2}g_{\mu
\nu }\left[ \left( \nabla \phi \right) ^{2}+2V\left( \phi \right) \right] , \\
T^{(\rm DM)}_{\mu\nu}&=&\frac{2\delta \left( \sqrt{-g}\mathcal{L}_{\rm DM}\right) }{%
\delta g^{\mu \nu }}.
\end{eqnarray}
}
Considering dark matter as a kind of perfect fluid, we can formulate the energy-momentum tensor in energy density and pressure terms as
\begin{equation}
T_{\mu }^{\nu }=\left[ -\rho ,p,p,p\right] .
\end{equation}%
In the simplest scenario, the energy density to pressure ratio is assumed to be constant, $\frac{\rho}{p}=\delta -1$, \cite{Li2012}. { Here, we posit that the dark energy of the universe is predominantly constituted by the potential of a scalar field in the form of \cite{Ratra1988} 
\begin{equation}
V\left( \phi \right) \simeq e^{-\sqrt{\frac{3}{2}\left( 1+\omega _{q}\right) 
}\phi },
\end{equation}%
which leads to a special solution such that the equation of state reads $P=\omega _{q}\rho .$ We now} determine the solution to the field equations for a static, spherically symmetric metric that takes the following form:
\begin{equation}
ds^{2}=-f\left( r\right) dt^{2}+\frac{1}{f\left( r\right) }%
dr^{2}+r^{2}d\theta ^{2}+r^{2}\sin ^{2}\theta d\varphi ^{2},
\end{equation}
where
\begin{equation}
 f(r)=1-\frac{2 m(r)}{r}.
\end{equation}
Regarding the electromagnetic field, we propose the following ansatz for the
Plebanski dual variables for the nonlinear electromagnetic field,%
\begin{equation}
P_{\mu \nu }=\frac{q}{r^{2}}\left( \delta _{\mu }^{0}\delta _{\nu
}^{1}-\delta _{\mu }^{1}\delta _{\nu }^{0}\right) ,
\end{equation}%
then the electromagnetic invariants read%
\begin{equation}
s=\frac{q^{2}}{2r^{4}}\text{ \ and }t=0.
\end{equation}%
When we insert these values into the $(0,0)$ component of the field equations (\ref{ef}), we find
\begin{equation}
\frac{dm}{dr}=\frac{q^{2}}{2r^{2}}-\frac{aq^{4}}{8r^{6}}-\frac{\alpha }{2r}-%
\frac{3\omega _{q}\sigma }{2r^{3\omega _{q}+1}}.
\end{equation}
After integrating this equation, we obtain the metric function of the Euler-Heisenberg black hole in the presence of quintessence matter and the PFDM background as follows:
\begin{equation}
 f(r)=1-\frac{2 M}{r}+\frac{q^2}{r^2}-\frac{a q^4}{20 r^6}-\frac{\sigma }{r^{3 \omega_{q} +1}}+\frac{\alpha }{r} \log \left(\frac{r}{{\lvert \alpha \rvert}}\right), \label{fun} 
\end{equation}
where $\sigma$, $\omega_{q}$ , $\alpha$ and { $q$} correspond to the quintessence matter normalization factor, quintessence state parameter, PFDM intensity and {the electric charge of black hole}, respectively. Before examining the thermal quantities, we would like to show qualitatively in graphs how quintessence matter and PFDM change the metric function. To do this, we first plot the metric function versus radius with three different quintessence state parameter values in Fig.~\ref{lafun}. 

\begin{figure}[htb!]
\begin{minipage}[t]{0.35\textwidth}
        \centering
        \subcaption{{$\omega_{q}=-0.35$}}
        \includegraphics[width=\textwidth]{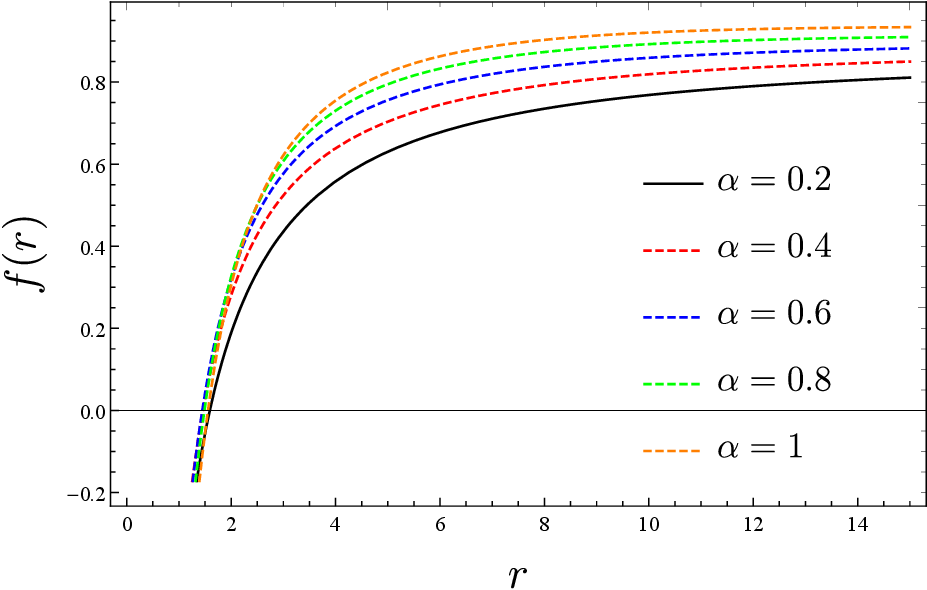}
         \label{fig:phbb}
\end{minipage}%
\begin{minipage}[t]{0.35\textwidth}
        \centering
        \subcaption{{$\omega_{q}=-0.65$}}
        \includegraphics[width=\textwidth]{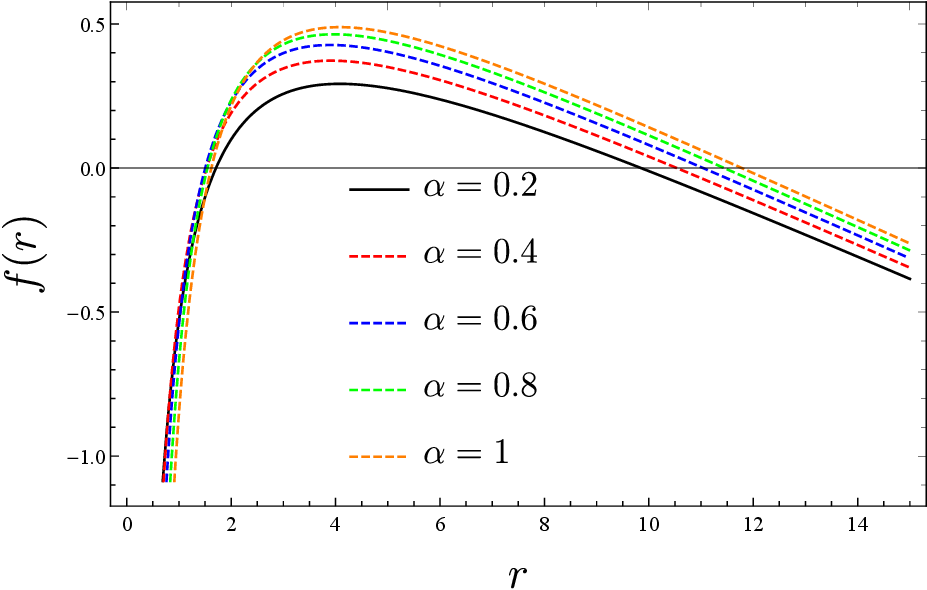}
       \label{fig:phcc}
   \end{minipage}%
\begin{minipage}[t]{0.35\textwidth}
        \centering
        \subcaption{{$\omega_{q}=-0.95$}}
        \includegraphics[width=\textwidth]{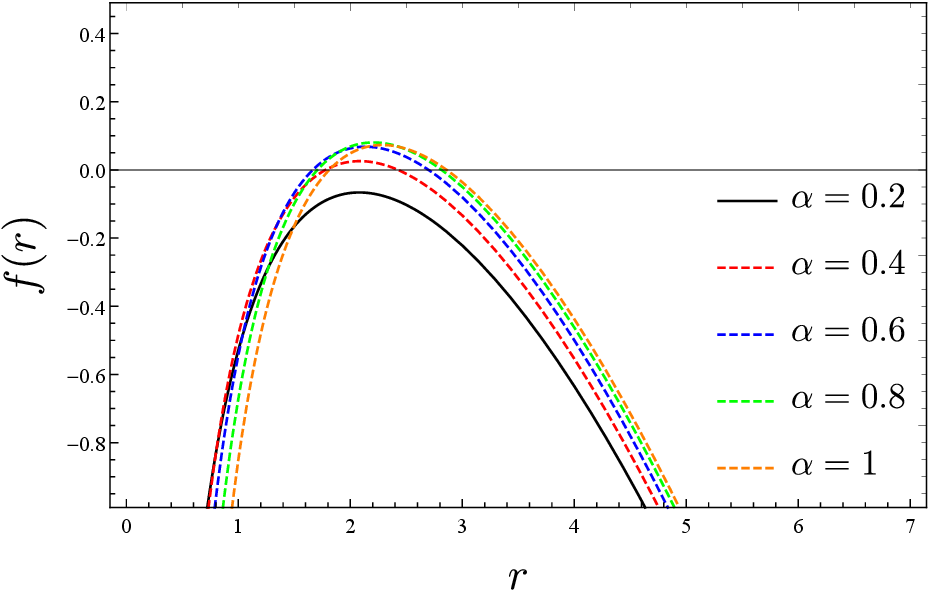}
         \label{fig:phdd}
   \end{minipage}
\caption{{{\footnotesize The impact of the PFDM on the metric function of the Euler-Heisenberg black hole for $M=1$, $q=0.5$, $a=0.1$, and $\sigma=0.1$.}}}
\label{laf}
\end{figure}

In each quintessence matter scenario, we consider different normalization factors with a constant impact on PFDM. We observe that for the $\omega_q=-0.35$ case, the metric function takes a zero value only at one radius. Here, we observe that for greater normalization constant values, this radius value receives larger values. These results differ from the other two scenarios. For example, for the $\omega_q=-0.65$ case, we note two different radii which lead to the metric function vanishing unless the normalization factor is less than a critical value. This behavior is also valid for smaller quintessence state parameter values, see the $\omega_q=-0.95$ case given in Fig.~\ref{fig:phd}. Next, we display the PFDM impact by taking the quintessence normalization parameter as a constant for three different scenario in Fig.~\ref{laf}.

In the $\omega_q=-0.35$ case, where the metric function takes a zero value at only one radius, increasing the PFDM intensity parameter decreases the radius. In the other scenario, in the $\omega_q=-0.65$, we observe two radii, and as the PFDM intensity increases, the interval between these radii increases.  In the third case, we see neatly that the presence of dark matter plays an essential role in the horizon radius. More precisely, for $\alpha=0.2$, the lapse function takes only negative values, meaning there is no event horizon, whereas for $\alpha=0.4$ there are two.

\newpage
\section{Thermodynamics} \label{sec:3}

\begin{figure}[htb!]
\begin{minipage}[t]{0.35\textwidth}
        \centering
        \subcaption{{$\omega_{q}=-0.35$ }}
        \includegraphics[width=\textwidth]{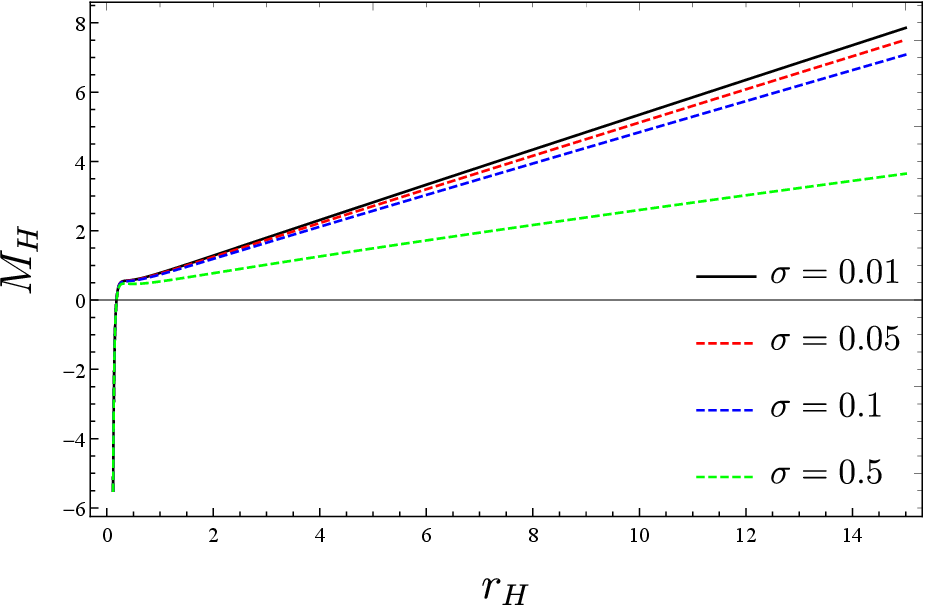}
         \label{fig:ma}
\end{minipage}%
\begin{minipage}[t]{0.35\textwidth}
        \centering
        \subcaption{{ $\omega_{q}=-0.65$}}
        \includegraphics[width=\textwidth]{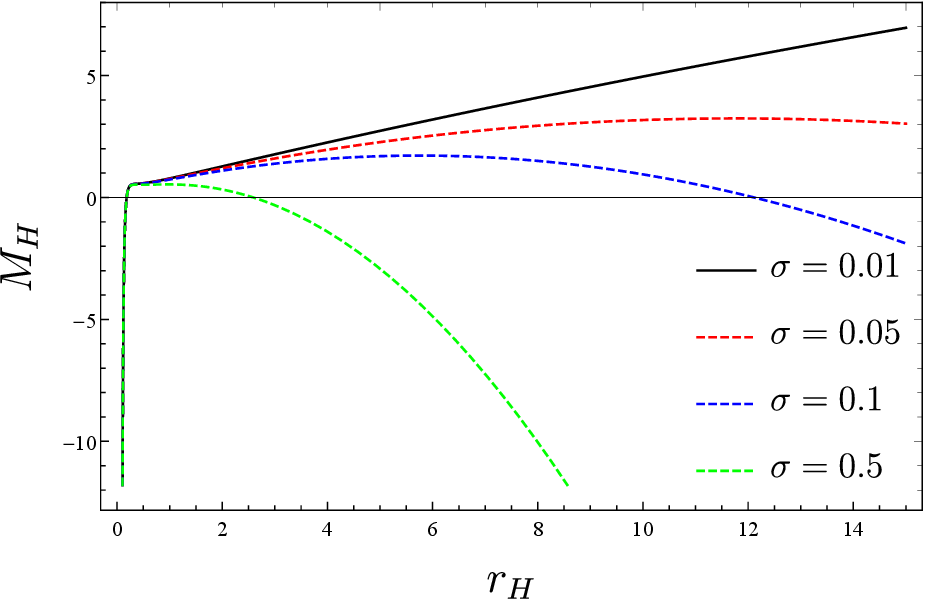}
       \label{fig:mb}
   \end{minipage}%
\begin{minipage}[t]{0.35\textwidth}
        \centering
        \subcaption{{$\omega_{q}=-0.95$}}
        \includegraphics[width=\textwidth]{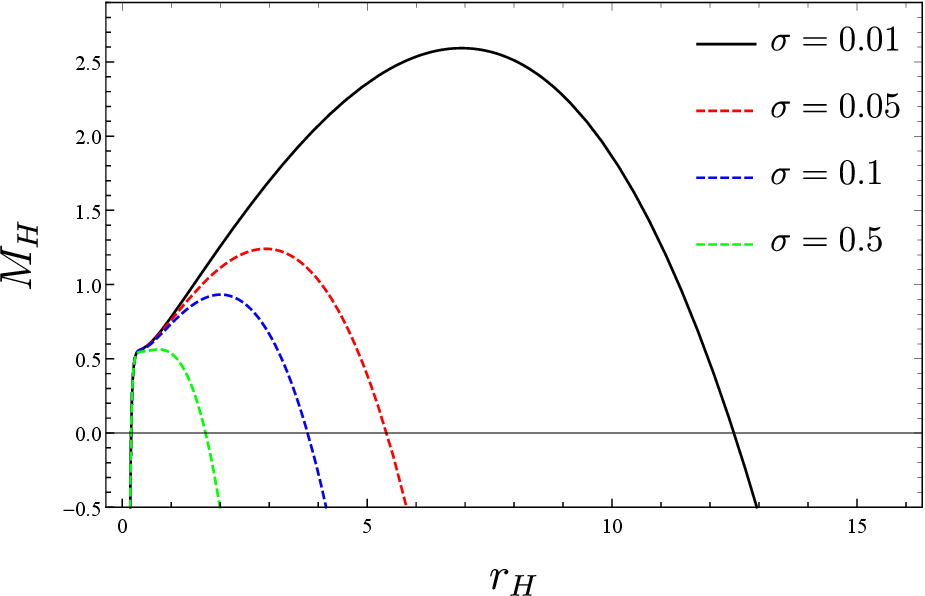}
         \label{fig:mc}
   \end{minipage}
\caption{{{\footnotesize The impact of the quintessence field on the mass of Euler-Heisenberg black hole for $q=0.5$, $a=0.1$ and $\alpha=0.2$.}}}
\label{massf}
\end{figure}

In this section, we begin to investigate the thermodynamics of the Euler-Heisenberg black hole by determining the black hole mass function at its horizon. To do this, we use the metric function as zero at the horizon and thus express the mass function in the following form
\begin{equation}
M_{H}={\frac{r_{H}}{2}}\left( 1+\frac{q^{2}}{r_{H}^{2}}-\frac{aq^{4}}{%
2r_{H}^{6}}-\frac{\sigma }{r_{H}^{1+3\omega _{q}}}+\frac{\alpha }{r_{H}}%
\log \left( \frac{r_{H}}{{\lvert \alpha \rvert} }\right) \right). \label{mass}
\end{equation}
We then graphically examine the effects of quintessential matter and PFDM on the mass function. Fig.~\ref{massf} reflects the impact of the quintessence matter. 

\begin{figure}[htb!]
\begin{minipage}[t]{0.35\textwidth}
        \centering
        \subcaption{{$\omega_{q}=-0.35$}}
        \includegraphics[width=\textwidth]{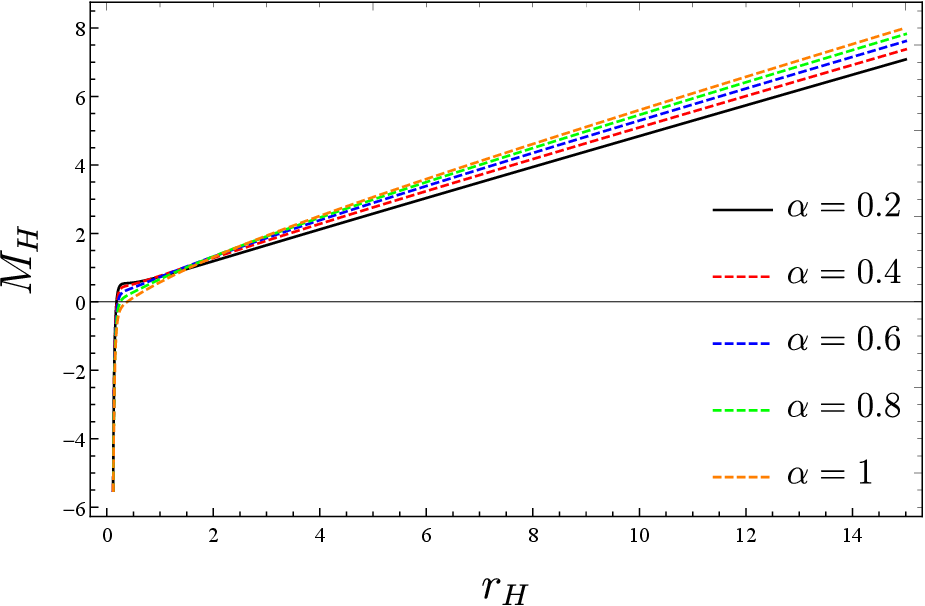}
         \label{fig:mbb}
\end{minipage}%
\begin{minipage}[t]{0.35\textwidth}
        \centering
        \subcaption{{$\omega_{q}=-0.65$}}
        \includegraphics[width=\textwidth]{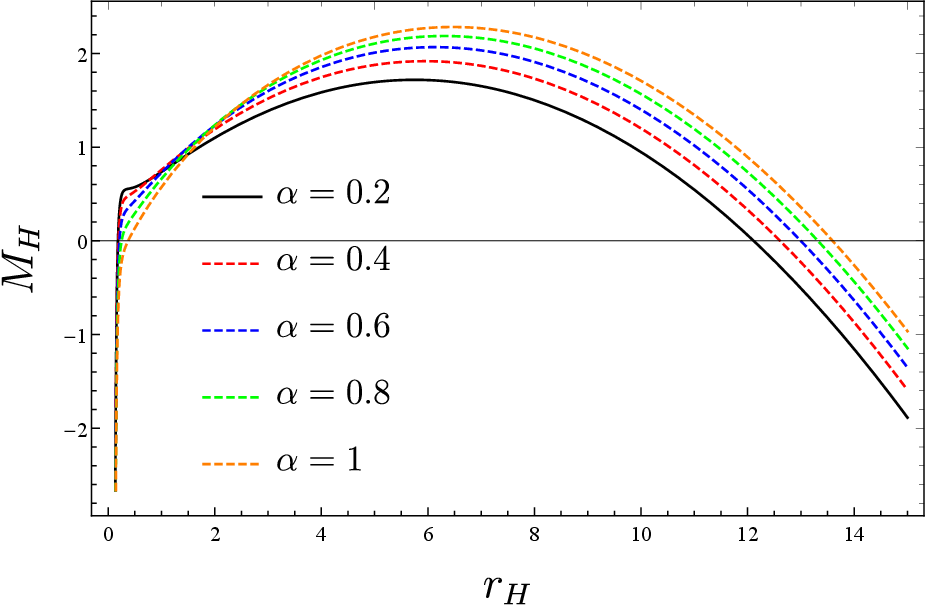}
       \label{fig:mcc}
   \end{minipage}%
\begin{minipage}[t]{0.35\textwidth}
        \centering
        \subcaption{{$\omega_{q}=-0.95$}}
        \includegraphics[width=\textwidth]{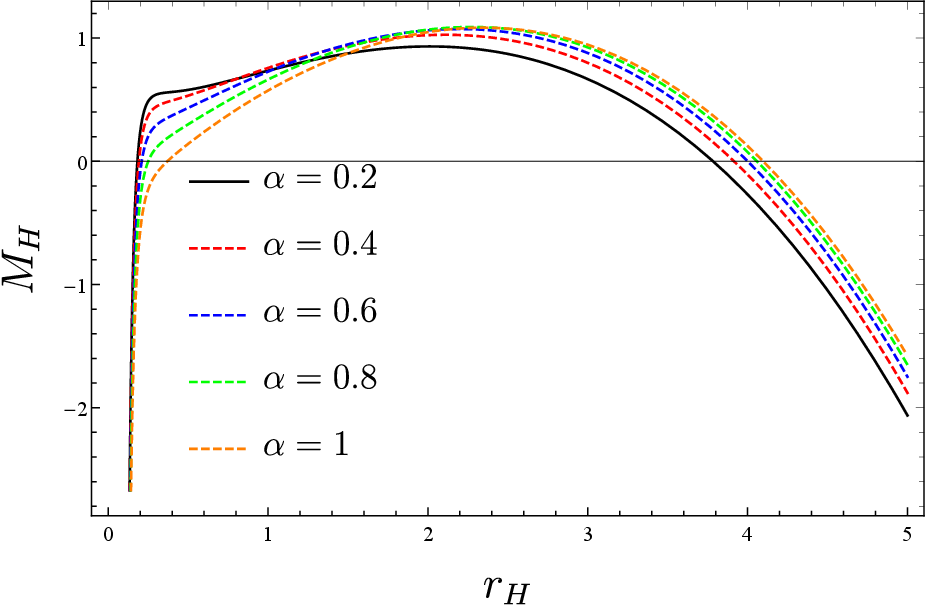}
         \label{fig:mdd}
   \end{minipage}
\caption{{{\footnotesize The impact of the dark matter on the mass of Euler-Heisenberg black hole for $q=0.5$, $a=0.1$ and $\sigma=0.1$.}}}
\label{maff}
\end{figure}

Fig.~\ref{fig:ma} shows that the mass function tends to increase monotonically and that this change is relatively smaller for relatively large values of the quintessence normalization parameter. In the cases of $\omega_q=-0.65$ and $\omega_q=-0.95$, we find that the black hole mass is physically limited by the upper horizon limits, in contrast to the previous scenario. This suggests the existence of a phase transition, which will be discussed later using the heat capacity function. We also note that in both cases the mass function reaches its highest values at larger radii when the quintessence normalization parameter has smaller values. Next, we explore the PFDM impact with Fig.~\ref{maff}. 

We observe that the dark matter affects the mass function but does not change its characteristic properties.  For example, in the $\omega_q=-0.35$ scenario, the mass function increases monotonically, which is relatively large for the denser PFDM. In the other two cases, we note that for greater intensity parameters, the physical range of the horizon enlarges while the highest values of mass function increase. 

\begin{figure}[htb!]
\begin{minipage}[t]{0.35\textwidth}
        \centering
        \subcaption{{$\omega_{q}=-0.35$}}
        \includegraphics[width=\textwidth]{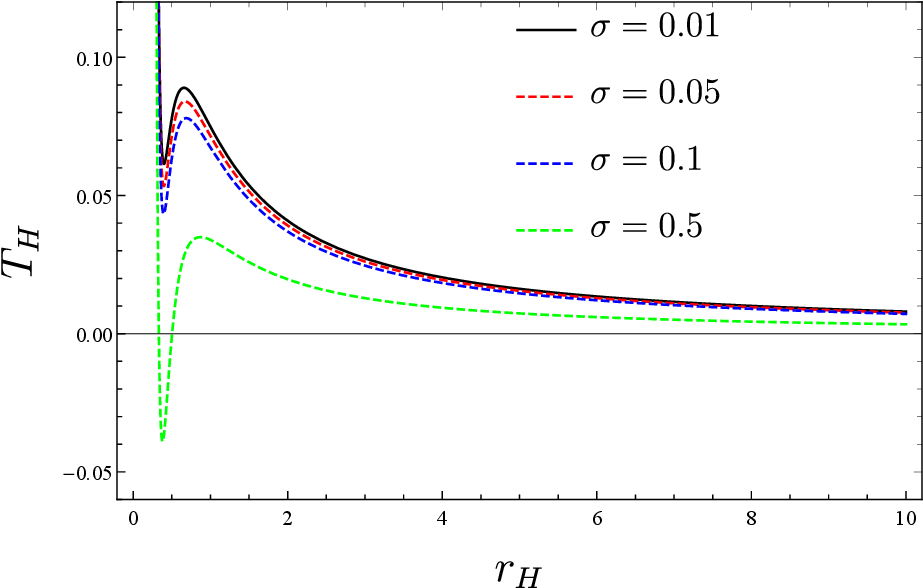}
         \label{fig:ta}
\end{minipage}%
\begin{minipage}[t]{0.35\textwidth}
        \centering
        \subcaption{{$\omega_{q}=-0.65$}}
        \includegraphics[width=\textwidth]{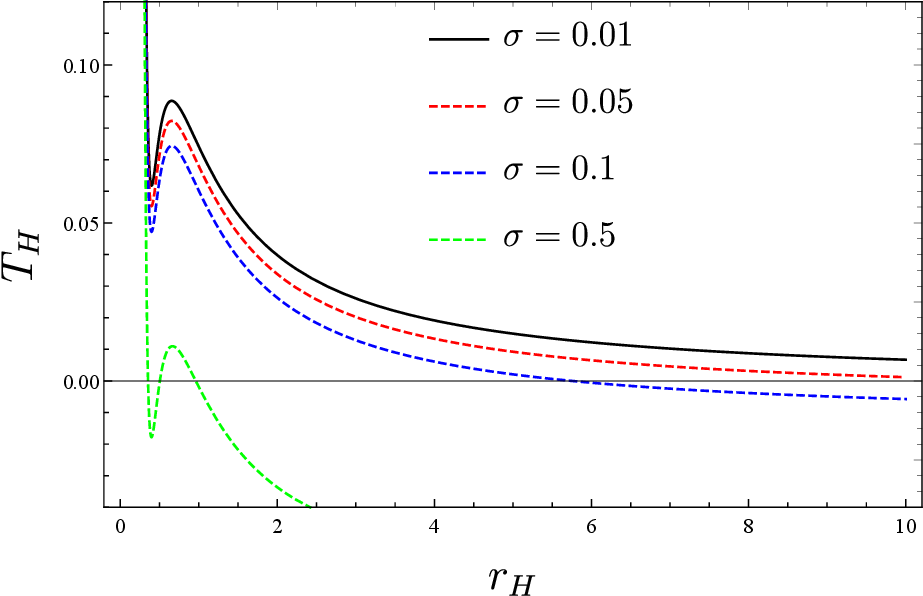}
       \label{fig:tb}
   \end{minipage}%
\begin{minipage}[t]{0.35\textwidth}
        \centering
        \subcaption{{$\omega_{q}=-0.95$}}
        \includegraphics[width=\textwidth]{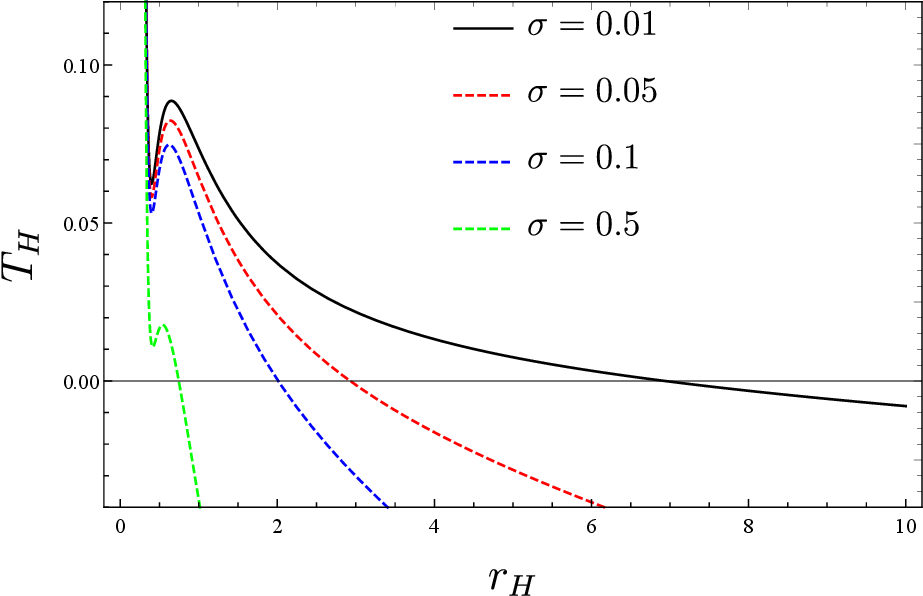}
         \label{fig:tc}
   \end{minipage}
\caption{{{\footnotesize The impact of the quintessence field on the temperature of Euler-Heisenberg black hole for $q=0.5$, $a=0.1$ and $\alpha=0.2$}}}
\label{tempf}
\end{figure}

Next, we employ 
\begin{equation}
T_{H}=\left. \frac{f^{\prime }\left( r\right) }{4\pi }\right\vert _{r=r_{H}},
\end{equation}%
to derive the Hawking temperature. We find
\begin{equation}
T_{H}=\frac{1}{4\pi r_{H}}\left( 1-\frac{q^{2}}{r_{H}^{2}}+\frac{aq^{4}}{%
4r_{H}^{6}}+\frac{3\sigma \omega _{q}}{%
r_{H}^{1+3\omega _{q}}}+\frac{\alpha }{r_{H}}\right).   \label{tem}
\end{equation}
We then demonstrate the plot of Hawking temperature versus horizon in Fig.~\ref{tempf}
and examine the effect of the quintessence field. 

In all scenarios, we notice that the black hole temperature first increases and then decreases. However, in the case of $\omega_q=-0.35$ the temperature tends to zero at infinity, while in the other cases, it equals zero at finite horizon radii. It should be noted that as the normalization coefficient of the quintessence increases, these event horizon radii decrease. We must emphasize that as the quintessence matter normalization coefficient increases, the temperature of the black hole remains at lower values. 

\begin{figure}[htb!]
\begin{minipage}[t]{0.35\textwidth}
        \centering
        \subcaption{{$\omega_{q}=-0.35$}}
        \includegraphics[width=\textwidth]{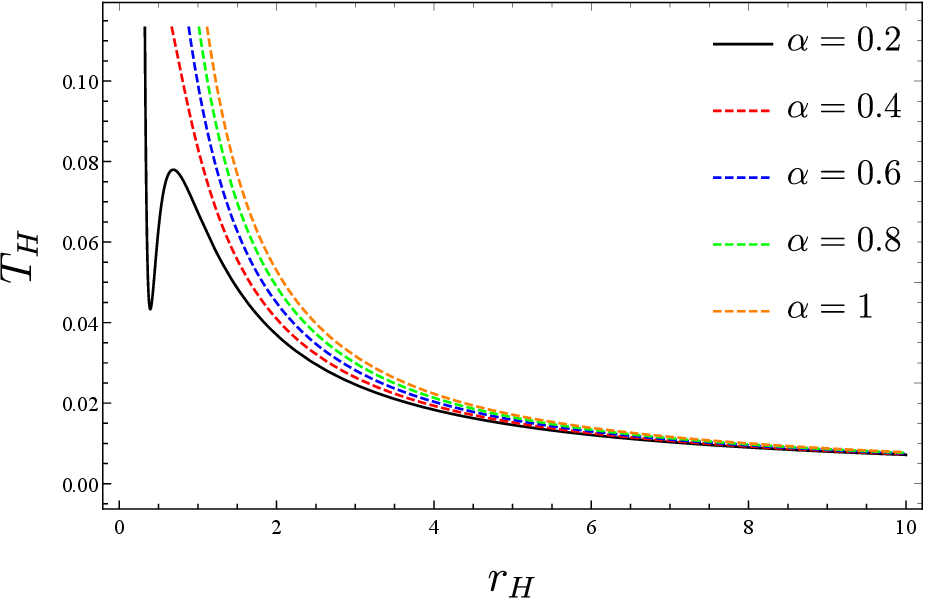}
         \label{fig:taa}
\end{minipage}%
\begin{minipage}[t]{0.35\textwidth}
        \centering
        \subcaption{{$\omega_{q}=-0.65$}}
        \includegraphics[width=\textwidth]{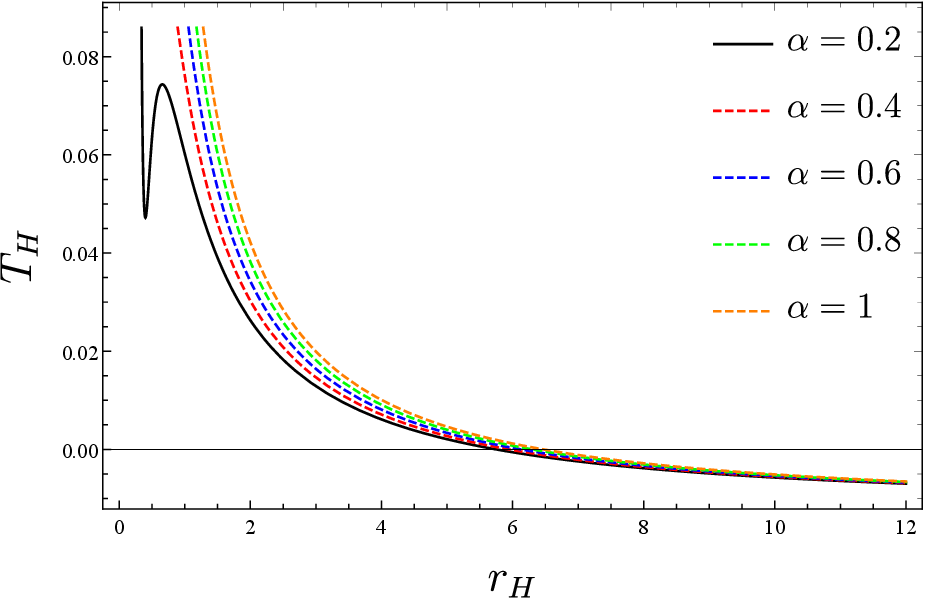}
       \label{fig:tbb}
   \end{minipage}%
\begin{minipage}[t]{0.35\textwidth}
        \centering
        \subcaption{{$\omega_{q}=-0.95$}}
        \includegraphics[width=\textwidth]{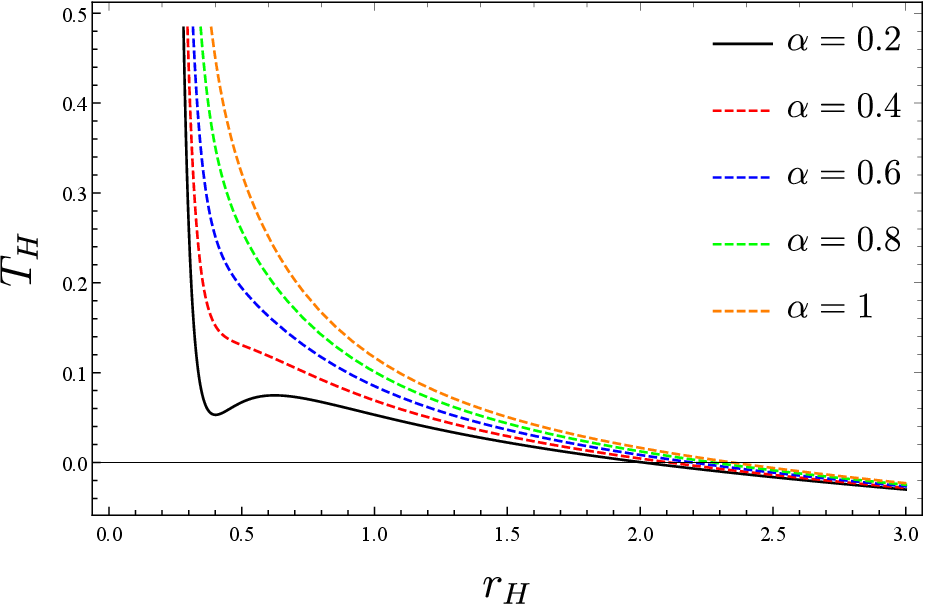}
         \label{fig:tcc}
   \end{minipage}
\caption{{{\footnotesize The impact of the dark matter on the temperature of Euler-Heisenberg black hole for $q=0.5$, $a=0.1$ and $\sigma=0.1$.}}}
\label{temf}
\end{figure}

Fig.~\ref{temf} shows the impact of the dark matter.

\newpage

We observe that the dark matter intensity has a crucial influence on the temperature behavior at relatively small horizon values. More specifically, in all scenarios, the initial decrease and subsequent increase behavior change to a monotonically decreasing behavior as the dark matter intensity increases. It is also worth noting that with higher dark matter intensity the Hawking temperature receives greater values in all cases. 

Then we study the entropy at the horizon via $S=\int \frac{dM}{T_H}$, and obtain it as follows:   
\begin{equation}
S=\pi r_{H}^{2}.
\end{equation}
This form reveals that the horizon entropy does not depend on either the quintessence matter or the dark matter fields. Therefore, we go through the next thermal quantity, the specific heat function, by employing $C=\frac{dM}{dT_H}$. We find
\begin{equation}
C=-2\pi r_{H}^{2}\frac{ 1-\frac{q^{2}}{r_{H}^{2}}+\frac{aq^{4}}{%
4r_{H}^{6}}+\frac{\alpha }{r_{H}}+\frac{3\sigma \omega _{q}}{%
r_{H}^{1+3\omega _{q}}} }{1-\frac{3q^{2}}{r_{H}^{2}}+\frac{7aq^{4}}{%
4r_{H}^{6}}+\frac{2\alpha }{r_{H}}+\frac{3\sigma \omega _{q}(3\omega _{q}+2)%
}{r_{H}^{1+3\omega _{q}}}}  \label{heat}
\end{equation}

Before, we depict the specific heat function we present important numerical values of the mass and the Hawking temperature functions in Tables \ref{tablecan1}, \ref{tablecan2} and \ref{tablecan3}. 
\begin{table}[htb!]
\centering%
\caption{{{\footnotesize ${T_H}_r$ and ${M}_{r}$  indicate the roots of temperature and mass, and ${T_H}_{tp1}$, ${T_H}_{tp2}$ and  ${M}_{tp}$ correspond to the turning points. Here, we take $a=0.1$, $q=0.5$ and $\alpha=0.2$.}}}
\begin{tabular}{c|c|c|c|c}
\hline
\rowcolor{lightgray} $\omega _{q}$ & $\sigma=0.01$ & $\sigma=0.05$ &
$\sigma=0.1$  &
$\sigma=0.5$ \\  \hline
\rowcolor{lightgray}{$-0.35$}  & $r_H$ & $r_H$ & $r_H$ & $r_H$   \\ \hline
\cellcolor{lightgray}${T_H}_r$  & - & - & - & $0.332460$, $0.503665$ \\ \hline
\cellcolor{lightgray}${T_H}_{tp1}$  & $0.398628$ & $0.396760$ & $0.394554$ & $0.380494$  \\
\hline
\cellcolor{lightgray}${T_H}_{tp2}$  & $0.657791$ & $0.670569$ & $0.687247$ & $0.870764$  \\
\hline
\cellcolor{lightgray}${M}_{r}$  & $0.182806$ & $0.182991$ & $0.183225$ & $0.185191$  \\
\hline
\cellcolor{lightgray}${M}_{tp}$  & - & - & -& $0.332460$, $0.503665$   \\
\hline
\cellcolor{lightgray}${r_H}_{phy}$  & $[0.182806, \infty)$ & $[0.182991, \infty)$ & 
$[0.183225, \infty)$ & $[0.185191, 0.332460]\, \bigcup \, [0.503665, \infty)$\\ \hline \hline
\end{tabular}%
\label{tablecan1}
\end{table}

\begin{table}[htb!]
\centering%
\caption{{{\footnotesize ${T_H}_r$, ${M}_{r1}$ and ${M}_{r2}$  indicate the roots of temperature and mass, and ${T_H}_{tp1}$, ${T_H}_{tp2}$ and  ${M}_{tp}$ correspond to the turning points. Here, we take $a=0.1$, $q=0.5$ and $\alpha=0.2$.}}}
\begin{tabular}{c|c|c|c|c}
\hline
\rowcolor{lightgray} $\omega _{q}$ & $\sigma=0.01$ & $\sigma=0.05$ &
$\sigma=0.1$  &
$\sigma=0.5$ \\  \hline
\rowcolor{lightgray}{$-0.65$}  & $r_H$ & $r_H$ & $r_H$ & $r_H$   \\ \hline
\cellcolor{lightgray}${T_H}_r$  &  $63.2962$ & $11.7786$ & $5.75925$ & $0.354091$,  $0.504547$,  $0.961937$ \\ \hline
\cellcolor{lightgray}${T_H}_{tp1}$  & $0.39909$ & $0.399007$ & $0.398904$ & $0.398089$  \\
\hline
\cellcolor{lightgray}${T_H}_{tp2}$  & $0.654875$ & $0.655709$ & $0.656756$ & $0.665360$  \\
\hline
\cellcolor{lightgray}${M}_{r1}$  & $0.182770$ & $0.182810$ & $0.182860$ & $0.183265$  \\
\hline
\cellcolor{lightgray}${M}_{r2}$  & $128.777$ & $24.3976$ & $12.1151$ & $2.59040$  \\
\hline
\cellcolor{lightgray}${M}_{tp}$  &  $63.2962$ & $11.7786$ & $5.75925$ & $0.354091$,  $0.504547$,  $0.961937$   \\
\hline
\cellcolor{lightgray}${r_H}_{phy}$  & $[0.182770, 63.2962]$ & $[0.182810, 11.7786]$ & 
$[0.182860, 5.75925]$ & \stackunder{$[0.183265, 0.354091]$}{$\bigcup \, [0.504547, 0.961937]$} \\ \hline
\hline
\end{tabular}%
\label{tablecan2}
\end{table}

\begin{table}[htb!]
\centering%
\caption{{{\footnotesize ${T_H}_r$, ${M}_{r1}$ and ${M}_{r2}$  indicate the roots of temperature and mass, and ${T_H}_{tp1}$, ${T_H}_{tp2}$ and  ${M}_{tp}$ correspond to the turning points. Here, we take $a=0.1$, $q=0.5$ and $\alpha=0.2$.}}}
\begin{tabular}{c|c|c|c|c}
\hline
\rowcolor{lightgray} $\omega _{q}$ & $\sigma=0.01$ & $\sigma=0.05$ &
$\sigma=0.1$  &
$\sigma=0.5$ \\  \hline
\rowcolor{lightgray}{$-0.95$}  & $r_H$ & $r_H$ & $r_H$ & $r_H$   \\ \hline
\cellcolor{lightgray}${T_H}_r$  &  $6.92959$ & $2.92693$ & $2.01074$ & $0.745429$  \\ \hline
\cellcolor{lightgray}${T_H}_{tp1}$  & $0.399338$ & $0.400265$ & $0.401472$ & $0.413965$  \\
\hline
\cellcolor{lightgray}${T_H}_{tp2}$  & $0.651193$ & $0.638197$ & $0.623624$ & $0.540601$  \\
\hline
\cellcolor{lightgray}${M}_{r1}$  & $0.182762$ & $0.182771$ & $0.182782$ & $0.182868$  \\
\hline
\cellcolor{lightgray}${M}_{r2}$  & $12.4878$ & $5.39633$ & $3.78417$ & $1.70234$  \\
\hline
\cellcolor{lightgray}${M}_{tp}$  &  $6.92959$ & $2.92693$ & $2.01074$ & $0.745429$   \\
\hline
\cellcolor{lightgray}${r_H}_{phy}$  & $[0.182762, 6.92959]$ & $[0.182771, 2.92693]$ & 
$[0.182782, 2.01074]$ & $[0.182808, 0.745429]$ \\ \hline \hline
\end{tabular}%
\label{tablecan3}
\end{table}

\begin{figure}[htb!]
\begin{minipage}[t]{0.35\textwidth}
        \centering
        \subcaption{{$\omega_{q}=-0.35$}}
        \includegraphics[width=\textwidth]{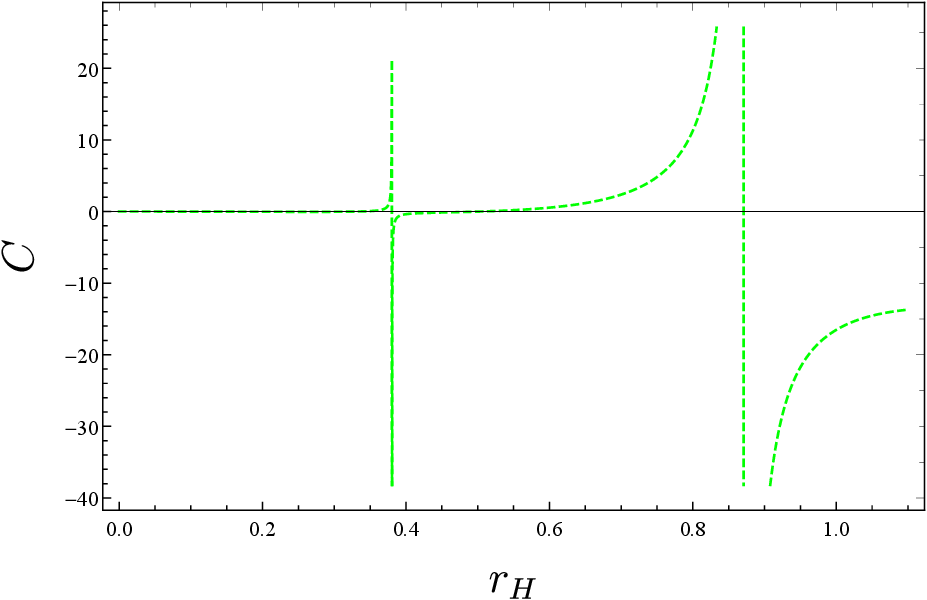}
         \label{cansp1a}
\end{minipage}%
\begin{minipage}[t]{0.35\textwidth}
        \centering
        \subcaption{{$\omega_{q}=-0.65$}}
        \includegraphics[width=\textwidth]{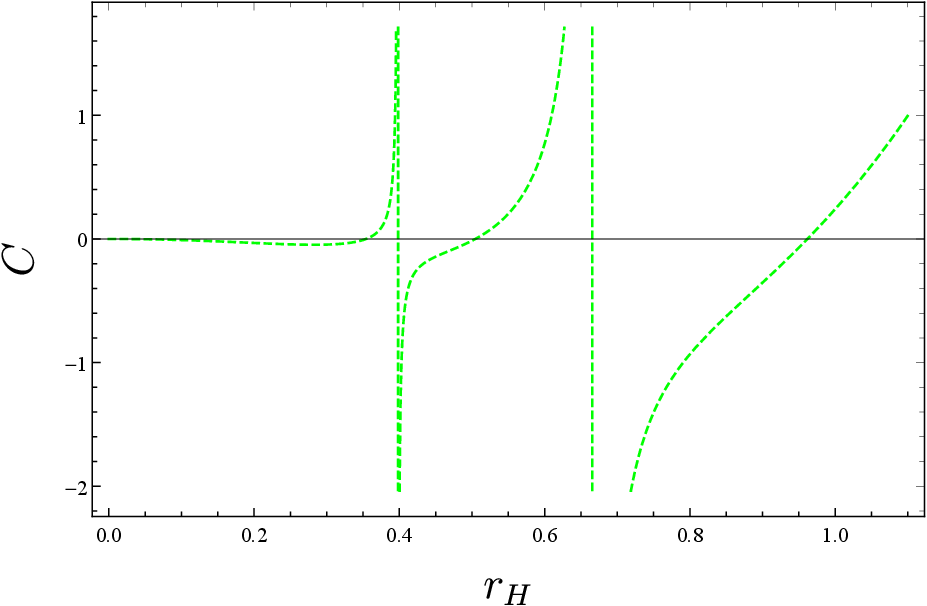}
       \label{cansp1b}
   \end{minipage}%
\begin{minipage}[t]{0.35\textwidth}
        \centering
        \subcaption{{$\omega_{q}=-0.95$}}
        \includegraphics[width=\textwidth]{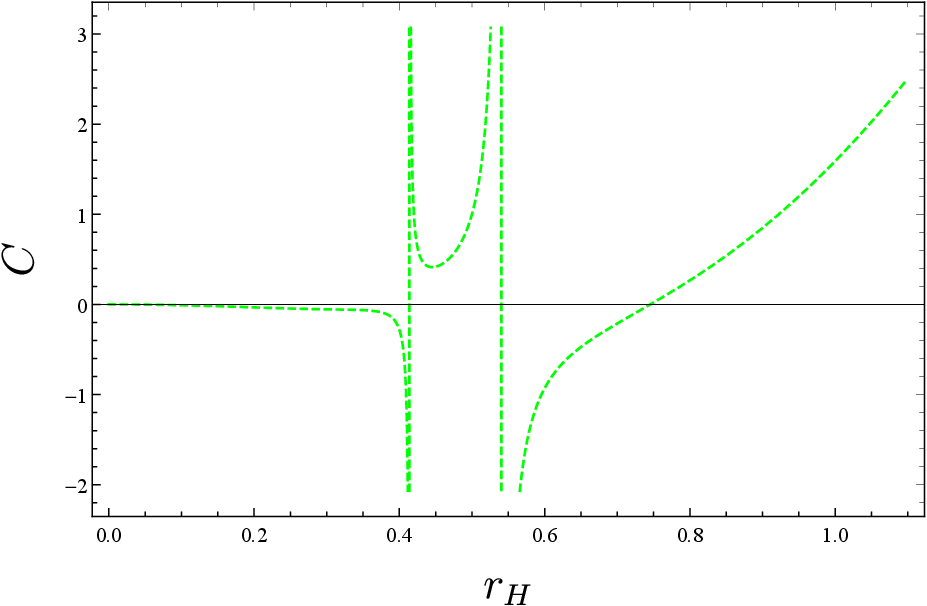}
         \label{cansp1c}
   \end{minipage}
\caption{{{\footnotesize Characteristic behavior of specific heat functions at relatively small event horizons for  $q=0.5$, $a=0.1$, $\alpha=0.2$ and $\sigma=0.5$.}}}
\label{cansp}
\end{figure}

In the $\omega_{q}=-0.35$ scenario, we observe that for sufficiently large $\sigma$ values, here $\sigma=0.5$, the physically meaningful horizon radius differs from the other cases. As shown in Fig.~\ref{cansp1a}, the black hole is unstable in the horizon range $[0.185191, 0.332460]$. We find that in this case, there is a remnant mass equal to $0.475903$ at $r_H=0.332460$. Then, in the horizon interval  $[0.503665,0870764)$ a stable black hole can appear, while in the $(0.870764,\infty)$ horizon an unstable one. It is worth noting that in this case the singularity of the specific heat function at $r_H=0.380494$ does not play a role, since it is not in the physically meaningful region, unlike the other one at $r_H=0.870764$. In the case of $\omega_{q}=-0.65$ we see a similar scenario, except for a radius that limits the event horizon from above. In this case we find a remnant mass with a value of $0.536008$ at $r_H=0.961937$. Interestingly, we find that $\omega_q=-0.95$ case is not similar to the previous ones discussed above. In this scenario, the black hole first has an unstable phase between $[0.182808, 0.413965)$, then a stable phase between $(0.413965, 0.961937)$.

\begin{figure}[htb!]
\begin{minipage}[t]{0.35\textwidth}
        \centering
        \subcaption{{$\omega_{q}=-0.35$}}
        \includegraphics[width=\textwidth]{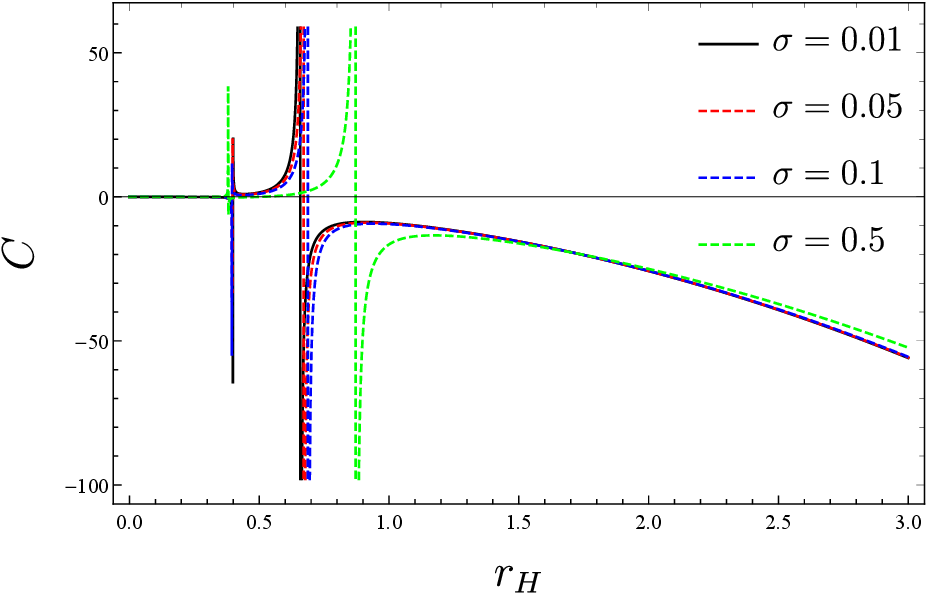}
         \label{fig:ca}
\end{minipage}%
\begin{minipage}[t]{0.35\textwidth}
        \centering
        \subcaption{{$\omega_{q}=-0.65$}}
        \includegraphics[width=\textwidth]{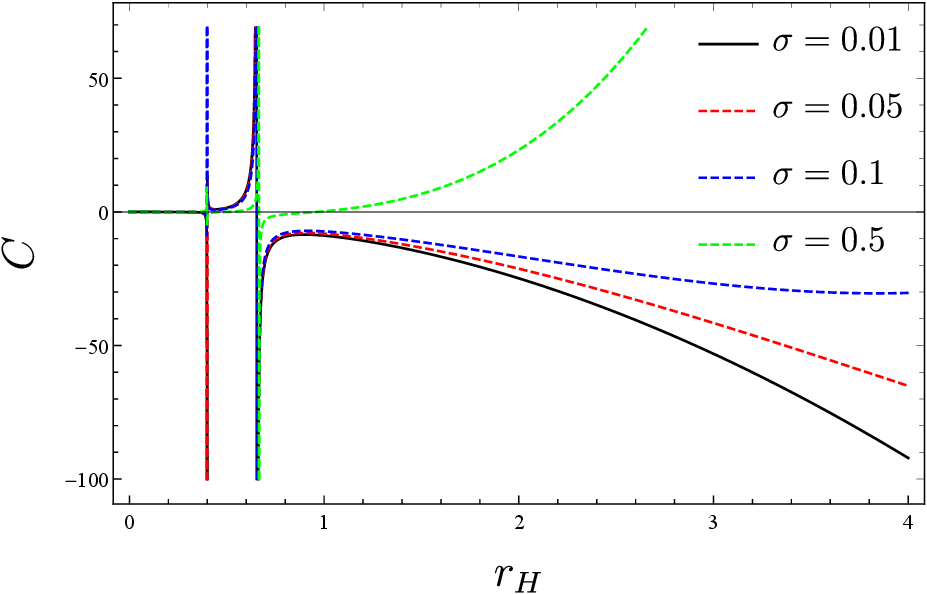}
       \label{fig:cb}
   \end{minipage}%
\begin{minipage}[t]{0.35\textwidth}
        \centering
        \subcaption{{$\omega_{q}=-0.95$}}
        \includegraphics[width=\textwidth]{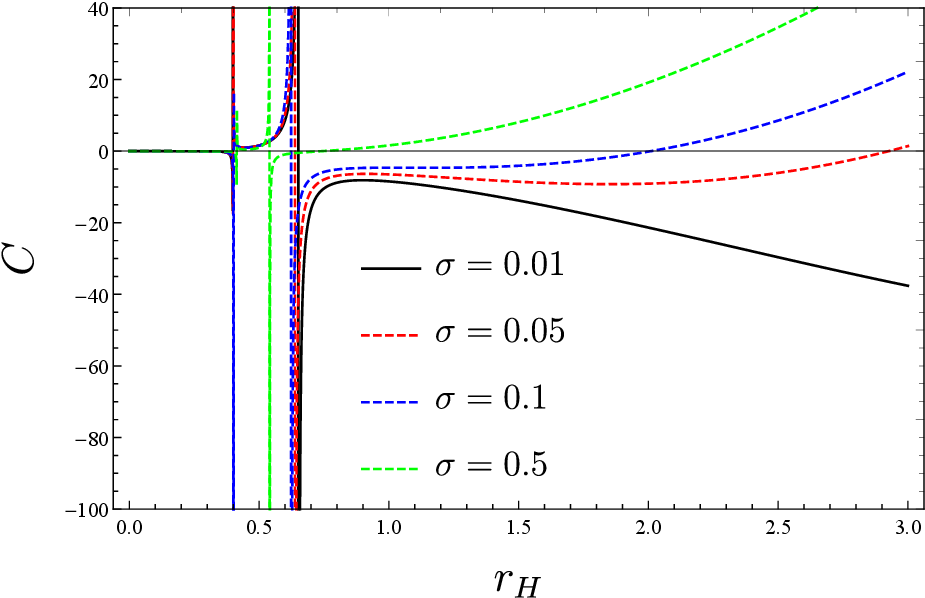}
         \label{fig:cc}
   \end{minipage}
\caption{{{\footnotesize The impact of the quintessence field on the e heat capacity function of Euler-Heisenberg black hole for $q=0.5$, $a=0.1$ and $\alpha=0.2$}}}
\label{heatf}
\end{figure}

\begin{figure}[htb!]
\begin{minipage}[t]{0.35\textwidth}
        \centering
        \subcaption{{$\omega_{q}=-0.35$}}
        \includegraphics[width=\textwidth]{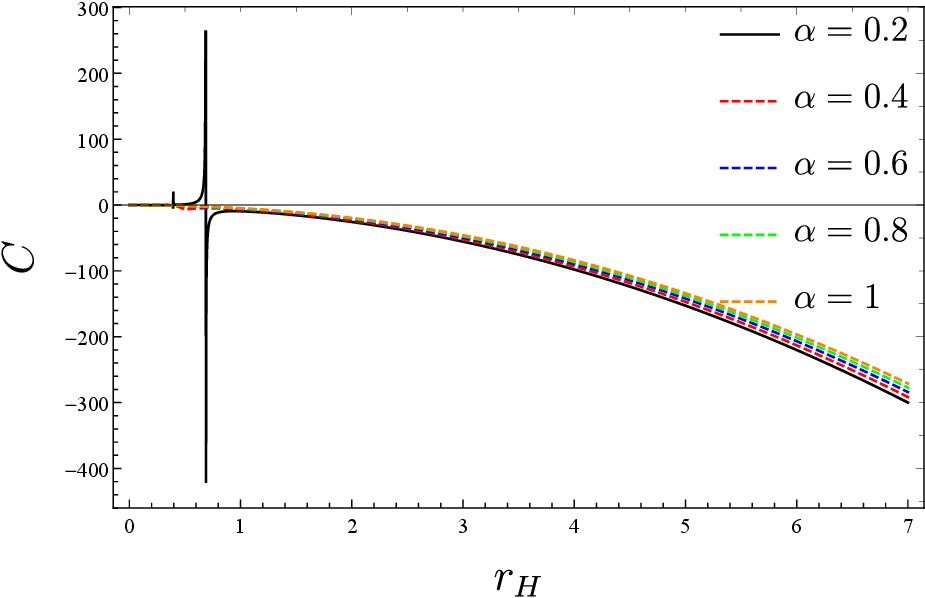}
         \label{fig:caa}
\end{minipage}%
\begin{minipage}[t]{0.35\textwidth}
        \centering
        \subcaption{{$\omega_{q}=-0.65$}}
        \includegraphics[width=\textwidth]{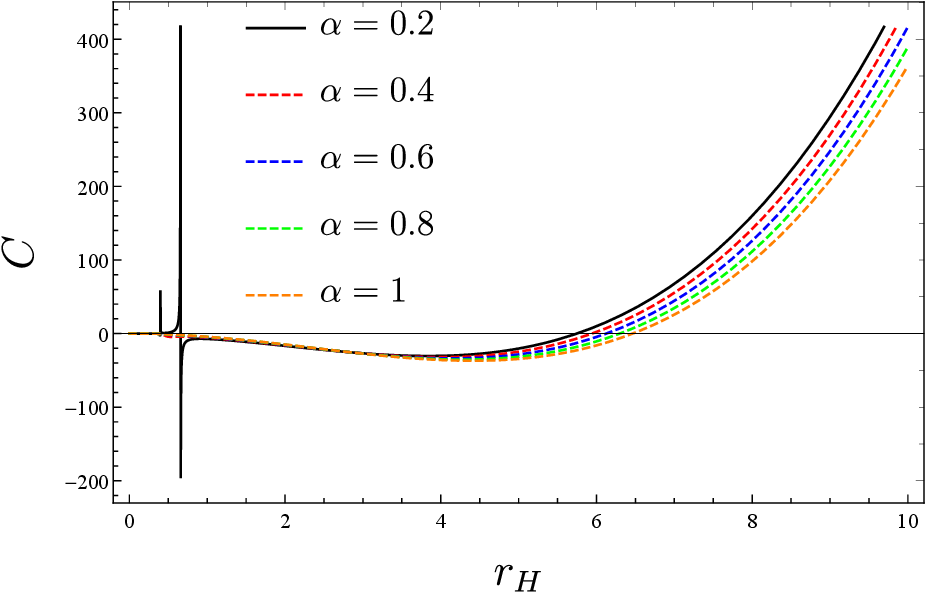}
       \label{fig:cbb}
   \end{minipage}%
\begin{minipage}[t]{0.35\textwidth}
        \centering
        \subcaption{{$\omega_{q}=-0.95$}}
        \includegraphics[width=\textwidth]{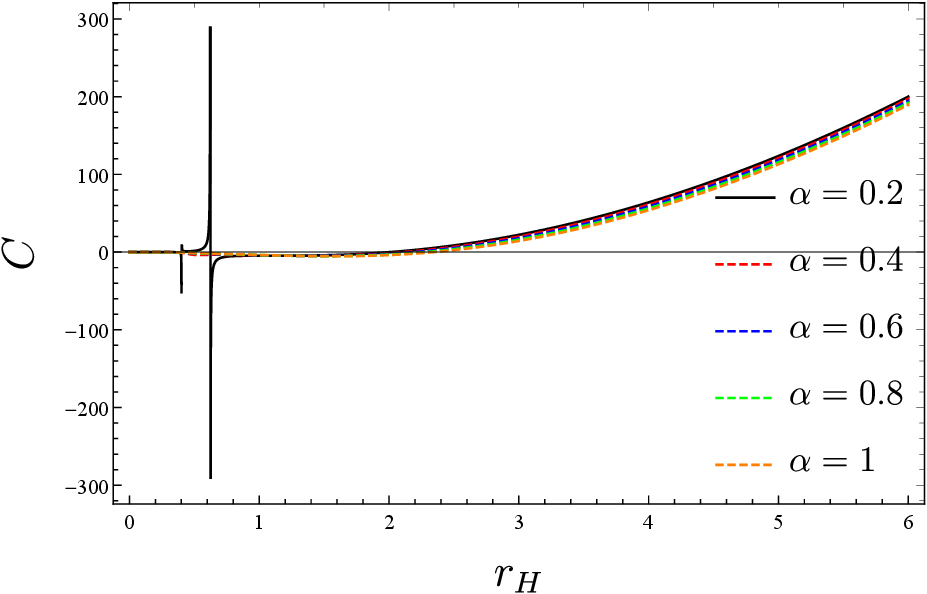}
         \label{fig:ccc}
   \end{minipage}
\caption{{{\footnotesize The impact of the dark matter on the heat capacity function of Euler-Heisenberg black hole for $q=0.5$, $a=0.1$ and $\sigma=0.1$}}}
\label{heatff}
\end{figure}

We then present the whole set of specific heat plots in Figs. \ref{heatf} and \ref{heatff}. It is worth noting that  Fig.~\ref{heatf} shows the accuracy of the physically meaningful ranges and limitations discussed above. 

\newpage
Fig.~\ref{heatff} shows that for relatively small dark matter intensities, the black hole has unstable, stable, and again unstable phases, respectively. However, for sufficiently large dark matter intensities, there is no phase transition and the black hole remains solely in an unstable phase. 

\begin{figure}[htb!]
\centering
\begin{minipage}[t]{0.43\textwidth}
        \centering
        \includegraphics[width=\textwidth]{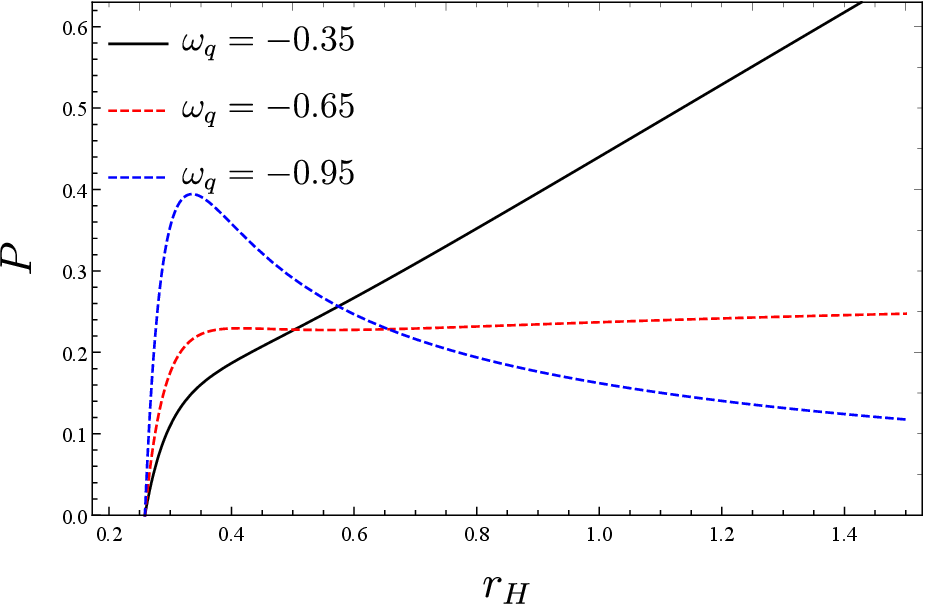}
\end{minipage}
\caption{{{\footnotesize The impact of the quintessence field on the pressure of Euler-Heisenberg black hole for $T=1$, $q=0.1$, $\alpha=0.2$ and $a=0.1$ }}}
\label{press}
\end{figure}

Next, using Eq. \eqref{tem} with $\sigma=-8\pi P$, we derive the equation of state 
\begin{equation}
P=\frac{r^{3\omega _{q}+1}}{24\pi \omega _{q}}\left( 1-\frac{q^{2}}{r_{H}^{2}%
}+\frac{aq^{4}}{4r_{H}^{6}}+\frac{\alpha }{r_{H}}-4\pi r_{H}T\right), 
\label{pre}
\end{equation}
where $P$ is the pressure. To demonstrate the effect of the dark energy we plot Fig.~\ref{press}.  

\begin{figure}[htb!]
\begin{minipage}[t]{0.35\textwidth}
        \centering
        \subcaption{{$\omega_{q}=-0.35$}}
        \includegraphics[width=\textwidth]{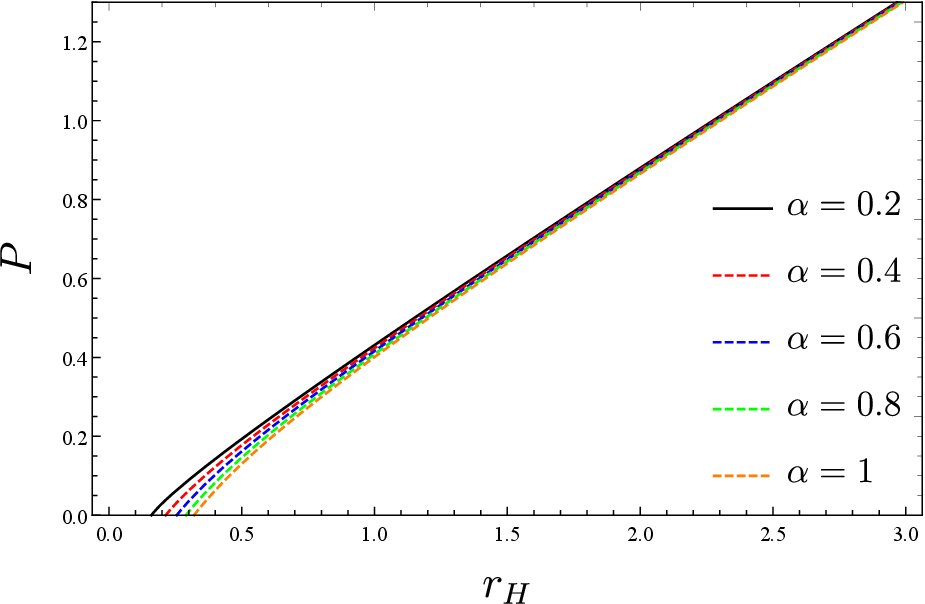}
         \label{fig:wa}
\end{minipage}%
\begin{minipage}[t]{0.35\textwidth}
        \centering
        \subcaption{{$\omega_{q}=-0.65$}}
        \includegraphics[width=\textwidth]{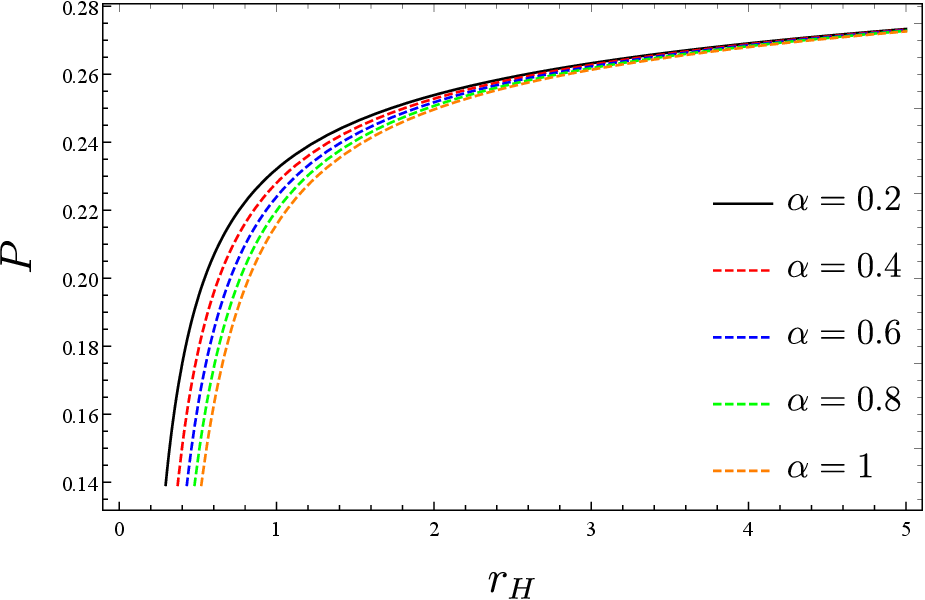}
       \label{fig:wb}
   \end{minipage}%
\begin{minipage}[t]{0.35\textwidth}
        \centering
        \subcaption{{$\omega_{q}=-0.95$}}
        \includegraphics[width=\textwidth]{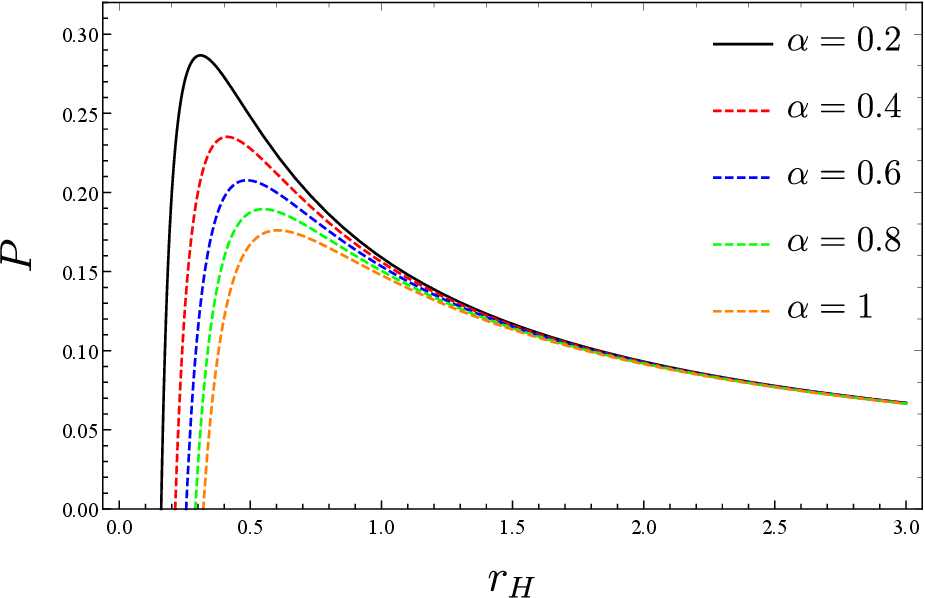}
         \label{fig:wc}
   \end{minipage}
\caption{{{\footnotesize The impact of the dark matter field on the pressure of Euler-Heisenberg black hole for $q=0.1$, $a=0.1$ }}}
\label{pressf}
\end{figure}

Fig.~\ref{press} shows that in the $\omega_q=-0.35$ scenario the pressure increases monotonically with increasing horizon radius. However, in other cases the pressure isotherms show different characteristics. For example, for $\omega_q=-0.65$ the pressure remains stable after an initial increase. In the case of $\omega_q=-0.95$ we observe a decrease after the initial increase and then a monotonous decrease. Then, in Fig.~\ref{pressf} we display the impact of dark matter. 

We find that dark matter also affects the pressure isotherms, especially at relatively small horizons.  In particular, with greater dark matter intensity, the pressure at the same radius in the $\omega_q=-0.35$ scenario takes on smaller values. In the $\omega_q=-0.65$ case, we see a similar behavior. However, in the $\omega_q=-0.95$ scenario, we observe a peak and the strength of these pressure peaks takes on smaller values at higher dark matter intensities.  

In this section, we finally focus on the critical points. To obtain them, we solve the following two equations:
\begin{equation}
\left. \frac{\partial P}{\partial r_{H}}=\frac{\partial ^{2}P}{\partial
r_{H}^{2}}\right \vert _{r_{H}=r_{c}}=0.
\end{equation}
We find 
\begin{equation}
T_{c}=\frac{\left( 3\omega _{q}-5\right) aq^{4}+4q^{2}r_{c}^{4}\left(
1-3\omega _{q}\right) +4r_{c}^{6}\left( 1+3\omega _{q}\right) +12\alpha
r_{c}^{5}\omega _{q}}{16\pi r_{c}^{7}(3\omega _{q}+2)},
\end{equation}%
\begin{equation}
P_{c}=\frac{r_{c}^{3\omega _{q}+1}\left( 1-\frac{3q^{2}}{r_{c}^{2}}+\frac{%
7aq^{4}}{4r_{c}^{6}}+2\alpha \right) }{24\pi \omega _{q}(3\omega _{q}+2)},
\end{equation}
and
\begin{equation}
\frac{r_{c}^{3\omega _{q}-1}}{24\pi \omega _{q}}\left( \Big(1+3\omega _{q}+\frac{%
6\alpha \omega _{q}}{r_{c}}\Big)+\frac{q^{2}(3-9\omega _{q})}{r_{c}^{2}}+\frac{%
7aq^{4}(3\omega _{q}-5)}{4r_{c}^{6}}\right) =0.\label{51}
\end{equation}
We now examine Eq. (\ref{51}) through numerical methods. We generate the critical point data and tabulate them in Table \ref{tabmm}. 
\begin{table}[htb!]
\centering%
\caption{{{\footnotesize Critical values for various PFDM parameters with $q=0.5$ and $a=0.1$.}}}
\small
\begin{tabular}{l|lll|lll|lll}
\hline\hline
\rowcolor{lightgray}\multirow{2}{*}{} & \multicolumn{3}{c|}{$\omega _{q}=-0.35
$} & \multicolumn{3}{c|}{$\omega _{q}=-0.65$} & \multicolumn{3}{c}{$\omega
_{q}=-0.95$} \\ \hline
\rowcolor{lightgray} $\alpha $ & $r_{c}$ & $T_{c}$ & $P_{c}$ & $r_{c}$ & $%
T_{c}$ & $P_{c}$ & $r_{c}$ & $T_{c}$ & $P_{c}$ \\ \hline
$0$ & \multicolumn{1}{|l}{0.456249} & 0.250427 & 0.0576791 & 
\multicolumn{1}{|l}{0.439973} & 0.246422 & 0.0567951 & \multicolumn{1}{|l}{
0.428451} & 0.237286 & 0.0548285 \\ 
$0.1$ & \multicolumn{1}{|l}{0.463858} & 0.208854 & 0.0396484 & 
\multicolumn{1}{|l}{0.450602} & 0.206674 & 0.0391574 & \multicolumn{1}{|l}{
0.440658} & 0.201482 & 0.0380114 \\ 
$0.2$ & \multicolumn{1}{|l}{0.472473} & 0.168705 & 0.0219448 & 
\multicolumn{1}{|l}{0.463513} & 0.167914 & 0.0217622 & \multicolumn{1}{|l}{
0.456191} & 0.165900 & 0.0213041 \\ 
$0.3$ & \multicolumn{1}{|l}{0.482397} & 0.130095 & 0.0046006 & 
\multicolumn{1}{|l}{0.479996} & 0.130052 & 0.0045904 & \multicolumn{1}{|l}{
0.477732} & 0.129929 & 0.0045613 \\ 
$0.4$ & \multicolumn{1}{|l}{0.494098} & 0.093168 & / & \multicolumn{1}{|l}{
0.503005} & 0.092751 & / & \multicolumn{1}{|l}{0.514610} & 0.091173 & / \\ 
\hline\hline
\end{tabular}
\label{tabm1}
\label{tabmm}
\end{table}
\normalsize

The current calculation demonstrates that the  PFDM parameter has a significant impact on the critical values. For example, an increase in $\alpha$, causes an increase in the critical radius, $r_{c}$. Concurrently, we notice that the critical temperature, $T_{c}$, and the critical pressure, $P_{c}$, both decrease as $\alpha$ increases.

\section{Black hole Shadow} \label{sec:4}
The black hole shadow holds immense significance in astrophysics as it offers a chance to validate Einstein's theory of general relativity. The shadow's dimensions and shape are governed by the curvature of spacetime surrounding the black hole, a key prediction of this theory. Through careful observation and analysis of the shadow, researchers can examine the theory's predictions, seeking any anomalies that might enhance our comprehension of the fundamental physical laws governing our Universe \cite{Zeng:2020dco, Zeng:2020vsj, Anacleto2021, pal1, Campos2022, Zeng:2021dlj, pal2, Anacleto2023, Ph24}. To study the shadow, our first step is to explore variations in the photon sphere. Utilizing Eq. (\ref{fun}) and following the conventional methodology outlined in \cite{Balendra}, we can numerically solve the equation to ascertain the precise location of the photon sphere. The equations governing the movement of photons in spacetime, dictated by the metric (\ref{fun}), can be derived from the Lagrangian formulation,
\begin{equation}
2\mathcal{L}=g_{\mu \nu }\dot{x}^{\mu }\dot{x}^{\nu },
\end{equation}
where the dot denotes differentiation with respect to an affine parameter. We now employ Eq. (\ref{fun}), as the line element,  and obtain the following form
\begin{equation}
2\mathcal{L}=-f\left( r\right) \dot{t}^{2}+\frac{1}{f\left( r\right) }\dot{r}%
^{2}+r^{2}\left( \dot{\theta}^{2}+\sin ^{2}\theta \dot{\varphi}^{2}\right) .
\end{equation}%
We then define the conjugate momenta 
\begin{equation}
p_{\mu }=\frac{\partial \mathcal{L}}{\partial \dot{x}^{\mu }}=g_{\mu \nu }\dot{x}^{\nu },
\end{equation}%
which yield two constants on motion%
\begin{eqnarray}
&&p_{t}=E=-f\left( r\right) \dot{t}, \\
&&p_{\varphi }=L=r^{2}\sin ^{2}\theta \dot{\varphi},
\end{eqnarray}
where $E$ represents the conserved energy quantity, while $L$ denotes the conserved angular momentum. To analyze the trajectories of photons orbiting the black hole, we employ the Hamilton-Jacobi method and utilize Carter's formulation of the geodesic equations \cite{Carter}. Now, we use the relativistic Hamilton-Jacobi equation, which reads stands in the following form
\begin{equation}
\frac{\partial }{\partial \tau }\mathcal{S}=-\frac{1}{2}g^{\mu \nu }\frac{%
\partial \mathcal{S}}{\partial x^{\mu }}\frac{\partial \mathcal{S}}{\partial
x^{\nu }},  \label{21}
\end{equation}
in which $\mathcal{S}$ and $\tau $ denote the Jacobi action and the affine parameter along the geodesic paths, respectively. We then suggest a separable solution for the Jacobi action in the following form:
\begin{equation}
\mathcal{S}=-Et+\mathcal{S}_{r}\left( r\right) +\mathcal{S}_{\theta }\left(
\theta \right) +L\varphi ,  \label{22}
\end{equation}
where $\mathcal{S}_{r}\left( r\right) $ and $\mathcal{S}_{\theta }\left(\theta \right) $ are functions that depend on the variables $r$ and $\theta$. By substituting Eq. (\ref{22}) into Eq. (\ref{21}) and employing the method of separating variables, we obtain the null geodesic equations that govern the motion of a photon around the Euler-Heisenberg black hole as follows:
\begin{equation}
-\frac{E^{2}}{f\left( r\right) }+f\left( r\right) \left( \frac{\partial 
\mathcal{S}_{r}}{\partial r}\right) ^{2}+\frac{1}{r^{2}}\left( \frac{%
\partial \mathcal{S}_{\theta }}{\partial \theta }\right) ^{2}+\frac{L^{2}}{%
r^{2}\sin ^{2}\theta }=0.\label{sep}
\end{equation}
Consequently, Eq. (\ref{sep}) can be expressed into the following two separate
equations:
\begin{eqnarray}
r^{4}f^{2}\left( r\right) \left( \frac{\partial \mathcal{S}_{r}}{\partial r}%
\right) ^{2}&=&r^{4}E^{2}-r^{2}f\left( r\right) \left( \mathcal{K}%
+L^{2}\right) ,  \label{sep1} \\
\left( \frac{\partial \mathcal{S}_{\theta }}{\partial \theta }\right) ^{2}&=&
\mathcal{K}-L^{2}\cot ^{2}\theta ,  \label{sep2}
\end{eqnarray}%
where $\mathcal{K}$ is Carter's the separation constant. Thus, we obtain the complete null geodesic equations in the following form.  
\begin{eqnarray}
\dot{t}&=&\frac{E}{f\left( r\right) },  \label{se1} \\
\dot{\varphi}&=&\frac{L}{r^{2}\sin ^{2}\theta },  \label{se2} \\
r^{2}\dot{r}&=&\pm \sqrt{\mathcal{R}}=\pm \sqrt{r^{4}E^{2}-r^{2}f\left(
r\right) \left( \mathcal{K}+L^{2}\right) },  \label{se3} \\
r^{2}\dot{\theta}&=&\pm \sqrt{\Theta}=\pm \sqrt{\mathcal{K}%
-L^{2}\cot ^{2}\theta }.  \label{se4}
\end{eqnarray}
It is worth emphasizing that this set of geodesic equations characterizes the dynamical behavior of a photon orbiting the Euler-Heisenberg black hole surrounded by quintessence in the background of PFDM. To find the boundaries of the black hole shadow, one can rewrite the
radial null geodesic equation for the Euler-Heisenberg black hole as
\begin{equation}
\dot{r}^{2}+\mathcal{V}_{{\rm eff}}\left( r\right) =0,  \label{vpot}
\end{equation}
where $\mathcal{V}_{{\rm eff}}\left( r\right) $ is the effective potential for
radial photon's motion defined by,%
\begin{equation}
\mathcal{V}_{{\rm eff}}\left( r\right) =\left( \frac{\mathcal{K}+L^{2}}{r^{2}}%
\right) \left( 1-\frac{2M}{r}+\frac{q^{2}}{r^{2}}-\frac{aq^{4}}{20r^{6}}-%
\frac{\sigma }{r^{3\omega _{q}+1}}+\frac{\alpha }{r}\log \left( \frac{r}{%
{\lvert \alpha \rvert}}\right) \right) -E^{2}.
\end{equation}%
Then, we recall the conditions for unstable circular orbits:
\begin{equation}
\mathcal{V}_{{\rm eff}}\left( r_{{\rm ph}}\right) =0, \quad  \text{or} \quad \mathcal{R}\left(
r_{{\rm ph}}\right) =0,  \label{vpot1}
\end{equation}%
and 
\begin{equation}
\left. \frac{d}{dr}\mathcal{V}_{{\rm eff}}\left( r\right) \right\vert _{r=r_{{\rm ph}}}=0,
\quad  \text{or} \quad \left. \frac{d}{dr}\mathcal{R}\left( r\right) \right\vert
_{r=r_{{\rm ph}}}=0.  \label{vpot2}
\end{equation}%
Here, the photon sphere radius, $r_{{\rm ph}}$, refers to the specific value of $r$ at which the effective potential $\mathcal{V}_{{\rm eff}}\left( r_{{\rm ph}}\right) $ reaches its maximum value  ($\left. \frac{d^{2}}{dr^{2}}\mathcal{V}%
_{{\rm eff}}\left( r\right) \right\vert _{r=r_{{\rm ph}}}<0$). Based on the conditions introduced in Eq.~(\ref{vpot2}), the photon sphere radius can be determined by solving the following equation: 
\begin{equation}
\frac{2aq^{4}}{5r_{{\rm ph}}^{6}}+\frac{\alpha }{r_{{\rm ph}}}+\frac{6M}{r_{{\rm ph}}}+\frac{%
3\sigma (\omega _{q}+1)}{r_{{\rm ph}}^{1+3\omega _{q}}}-\frac{4q^{2}}{r_{{\rm ph}}^{2}}-%
\frac{3\alpha }{r_{{\rm ph}}}\log \left( \frac{r_{{\rm ph}}}{{\lvert \alpha \rvert} }\right) -2=0.\label{phsp}
\end{equation}
Now, we define two impact parameters $\xi $ and $\eta $ as,%
\begin{equation}
\xi =\frac{L}{E}, \quad \text{and} \quad \eta =\frac{\mathcal{K}}{E^{2}}.
\label{18}
\end{equation}%
Recalling the condition for unstable circular orbits, given in Eq. (\ref{vpot1}), and
introducing two impact parameters  $\xi $ and $\eta $,  we execute several algebraic manipulations and obtain an expression for $\xi ^{2}+\eta $ as follows:
\begin{equation}
\xi ^{2}+\eta =\frac{r_{{\rm ph}}^{2}}{1-\frac{2M}{r_{{\rm ph}}}+\frac{q^{2}}{r_{{\rm ph}}^{2}}%
-\frac{aq^{4}}{20r_{{\rm ph}}^{6}}-\frac{\sigma }{r_{{\rm ph}}^{3\omega _{q}+1}}+\frac{%
\alpha }{r_{{\rm ph}}}\log \left( \frac{r_{{\rm ph}}}{{\lvert \alpha \rvert} }\right) }.
\end{equation}
Since we notice that Eq.~\eqref{phsp} cannot be solved analytically, we decided to employ numerical methods to obtain solutions. We tabulate the photon sphere radius with the computed values of $\xi ^{2}+\eta $ in Tables \ref{taa} and \ref{tbb}, respectively.
\begin{table}[tbh]
\centering
\caption{{{\footnotesize The values of the photon radius and $\eta +\zeta^{2}$ for different values of dark matter intensity parameter and $\omega _{q}$ with $\sigma =0.1$, $a=0.1$, $q=0.5$ and $M=1$. }}}
\begin{tabular}{l|ll|ll|ll}
\hline\hline
\rowcolor{lightgray}\multirow{2}{*}{} & \multicolumn{2}{l|}{$\omega
_{q}=-0.35$} & \multicolumn{2}{l|}{$\omega _{q}=-0.65$} & \multicolumn{2}{l}{%
$\omega _{q}=-0.95$} \\ \hline
\rowcolor{lightgray}$\alpha $ & $r_{{\rm ph}}$ & $\xi ^{2}+\eta $ & $r_{{\rm ph}}$ & $%
\xi ^{2}+\eta $ & $r_{{\rm ph}}$ & $\xi ^{2}+\eta $ \\ \hline
$0.1$ & 2.63979 & 23.4421 & 2.74793 & 45.8711 & 2.46451 & / \\ 
$0.2$ & 2.38999 & 18.4211 & 2.44449 & 30.6887 & 2.22587 & / \\ 
$0.3$ & 2.24954 & 15.5766 & 2.27995 & 23.7410 & 2.09844 & / \\ 
$0.4$ & 2.17464 & 13.8827 & 2.19390 & 20.0809 & 2.03507 & 165.209 \\ 
$0.5$ & 2.14234 & 12.8698 & 2.15688 & 18.0465 & 2.01228 & 82.2836 \\ 
\hline\hline
\end{tabular}
\label{taa}
\end{table}

\begin{table}[tbh]
 \centering
 \caption{{{\footnotesize The values of the photon radius and $\eta +\zeta^{2}$ for different values of $\sigma $ and $\omega _{q}$ with $\alpha =0.2$, $a=0.1$, $q=0.5$ and $M=1$.}}}
\begin{tabular}{l|ll|ll|ll}
\hline\hline
\rowcolor{lightgray}\multirow{2}{*}{} & \multicolumn{2}{c}{$\omega _{q}=-0.35
$} & \multicolumn{2}{c}{$\omega _{q}=-0.65$} & \multicolumn{2}{c|}{$\omega
_{q}=-0.95$} \\ \hline
\rowcolor{lightgray} $\sigma $ & $r_{{\rm ph}}$ & $\xi ^{2}+\eta $ & $r_{{\rm ph}}$ & $%
\xi ^{2}+\eta $ & $r_{{\rm ph}}$ & $\xi ^{2}+\eta $ \\ \hline
$0.01$ & 2.17622 & 13.8180 & 2.17806 & 14.2562 & 2.16138 & 15.2263 \\ 
$0.04$ & 2.24301 & 15.1576 & 2.25399 & 17.5043 & 2.18170 & 25.6609 \\ 
$0.07$ & 2.31412 & 16.6810 & 2.34151 & 22.4233 & 2.20315 & 81.2867 \\ 
$0.1$ & 2.38999 & 18.4211 & 2.44449 & 30.6887 & 2.22587 & / \\ \hline\hline
\end{tabular}
\label{tbb}
\end{table}

\newpage
Table \ref{taa} shows how $\eta +\xi ^{2}$ varies with respect to the dark matter intensity and the quintessence state parameters.
From this table, we can see that for a constant value of $\omega _{q}$, the value of $\eta +\xi ^{2}$ decreases as $\alpha $
increases. On the other hand, if we fix the value of $\alpha $, the value of $\eta +\xi ^{2}$ increases as the quintessence parameter 
increases. Moreover, in Table \ref{tbb}, we present the computed values of the photon radius $r_{{\rm ph}}$ and $\eta +\xi ^{2}$ for different values of $\sigma $ and $\omega _{q}$. We observe that the impact of the quintessence normalization factor does not {mimic} the effect of the dark matter intensity parameter. More precisely, as $\sigma$ increases, the {photon radius} and the $\eta +\xi ^{2}$ increases. 

\begin{figure}[htb!]
\centering
\begin{minipage}[t]{0.4\textwidth}
        \centering
        \subcaption{{$\omega_{q}=-0.35$}}
        \includegraphics[width=\textwidth]{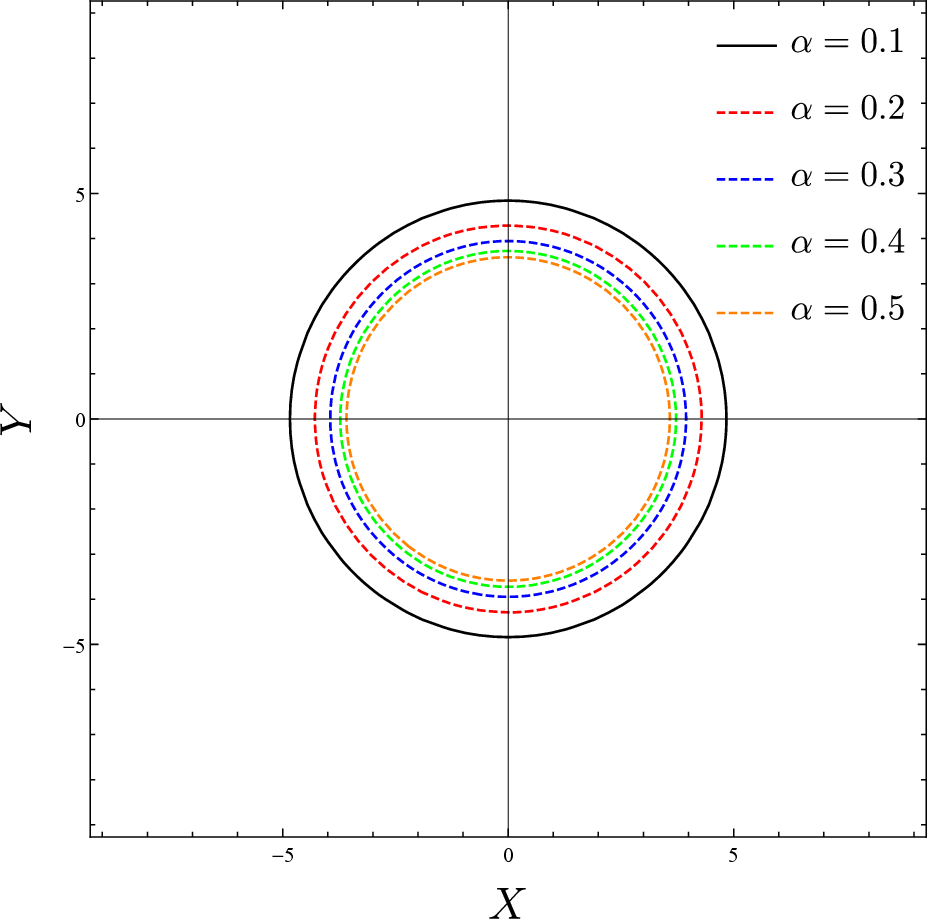}
         \label{fig:pha}
\end{minipage}%
\begin{minipage}[t]{0.4\textwidth}
        \centering
        \subcaption{{$\omega_{q}=-0.65$}}
        \includegraphics[width=\textwidth]{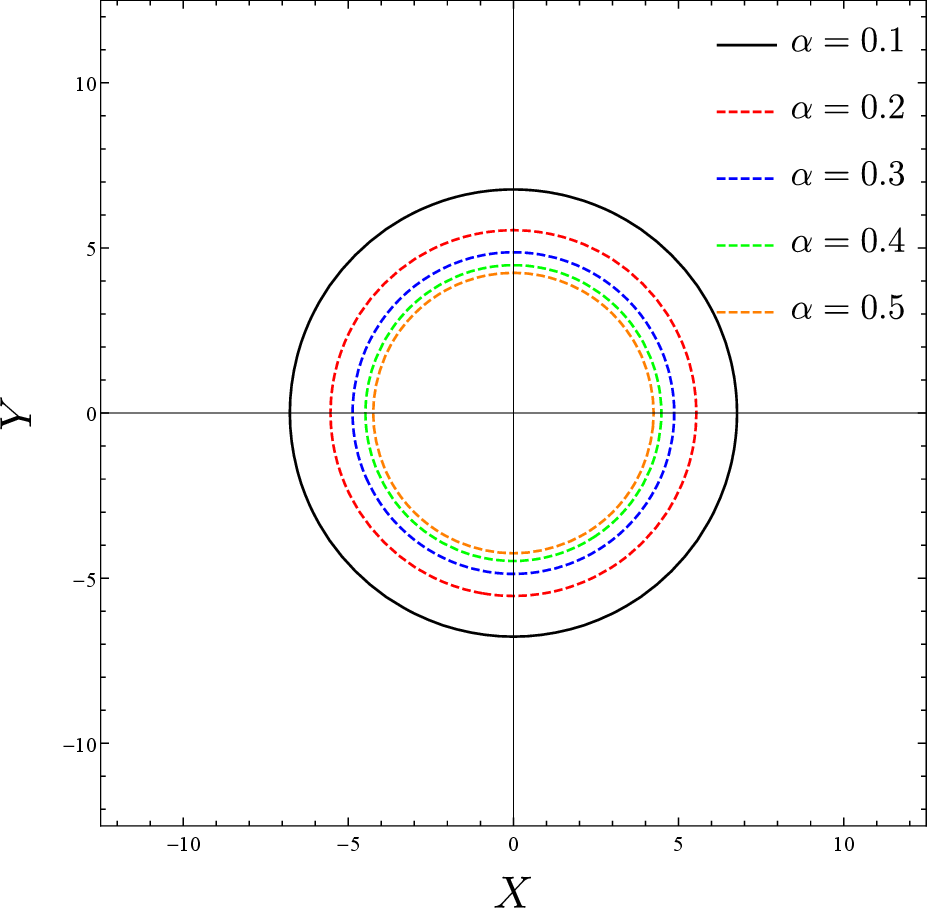}
       \label{phb}
   \end{minipage}%
\caption{{{\footnotesize Black hole shadow in the celestial plane ($X$-$Y$) for varying $\alpha$ for 2-values of quintessence $\omega_{q}=$ -0.35, and -0.65 }}}
\label{photsf}
\end{figure}

\begin{figure}[htb!]
\centering
\begin{minipage}[t]{0.4\textwidth}
        \centering
        \subcaption{{$\omega_{q}=-0.35$}}
        \includegraphics[width=\textwidth]{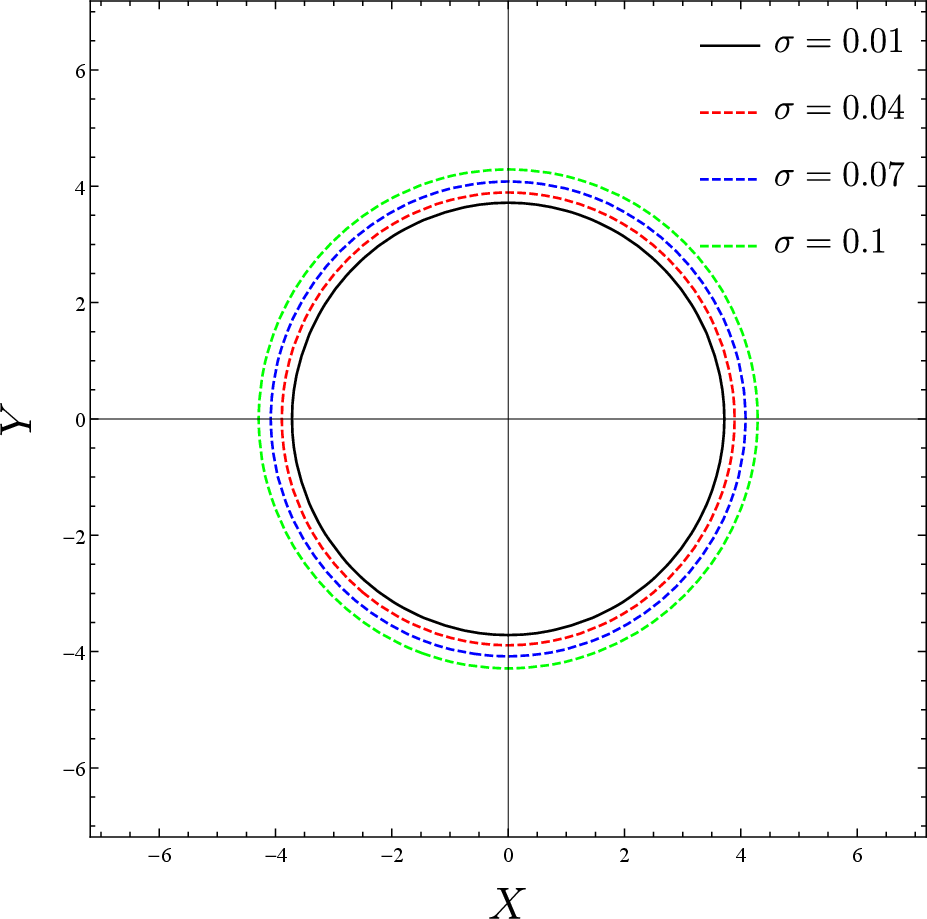}
         \label{fig:phad}
\end{minipage}%
\begin{minipage}[t]{0.4\textwidth}
        \centering
        \subcaption{{$\omega_{q}=-0.65$}}
        \includegraphics[width=\textwidth]{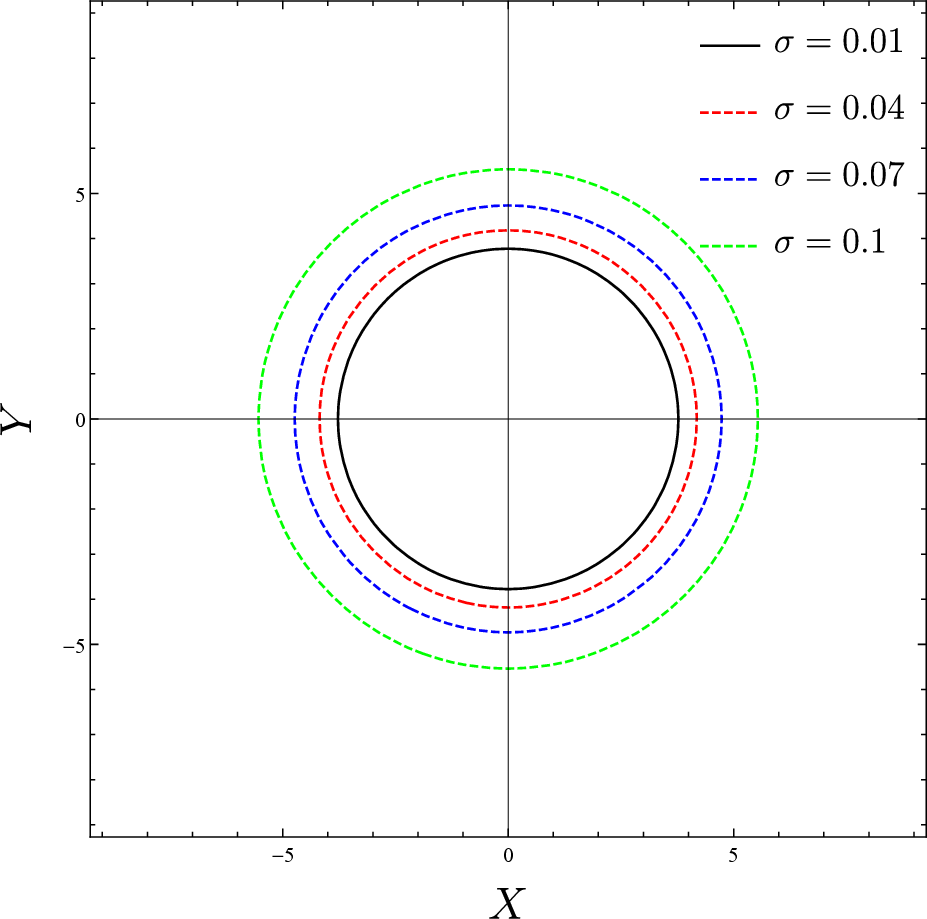}
       \label{fig:phbd}
   \end{minipage}%
\caption{{{\footnotesize Black hole shadow in the celestial plane ($X$-$Y$) for varying $\sigma$ for 2-values of quintessence $\omega_{q}=$ = -0.35 and -0.65 }}}
\label{photsfd}
\end{figure}

To visualize the shadow on the observer's frame correctly, one should employ the celestial coordinates $X$ and $Y$. In accordance with the methodology outlined in \cite{Balendra}, the celestial coordinates $X$ and $Y$ are defined as follows:
\begin{eqnarray}
X&=&\lim_{r_{o}\rightarrow \infty }\left( -r_{o}^{2}\sin \theta _{o}\frac{%
d\varphi }{dr}\right) ,  \label{XX} \\
Y&=&\lim_{r_{o}\rightarrow \infty }\left( r_{o}^{2}\frac{d\theta }{dr}\right) ,
\label{YY}
\end{eqnarray}
where $r_{o}$ represents the distance between the observer and the black hole, while $\theta _{o}$ denotes the angle formed by the black hole's rotation axis and the observer's line of sight. The values of $\frac{d\varphi }{dr}$ and $\frac{d\theta }{dr}$ can be determined by employing the geodesic equations, which are defined as:
{
\begin{eqnarray}
\frac{d\varphi }{dr}&=&\frac{\xi }{\sin ^{2}\theta _{o}\sqrt{%
r_{o}^{4}-r_{o}^{2}f\left( r_{o}\right) \left( \xi ^{2}+\eta \right) }}, \\
\frac{d\theta }{dr}&=&\frac{\sqrt{\eta -\xi ^{2}\cot ^{2}\theta _{o}}}{\sqrt{%
r_{o}^{4}-r_{o}^{2}f\left( r_{o}\right) \left( \xi ^{2}+\eta \right) }}.
\end{eqnarray}
}
By substituting the values obtained from the geodesic equations into the definitions of $X$ and $Y$, and then taking the limit as { $r_{o}\rightarrow
\infty $}, we arrive at the following expressions:
{
\begin{eqnarray}
X&=&-\frac{\zeta }{\sin \theta _{o}},\\
Y&=&\sqrt{\eta -\zeta ^{2}\cot ^{2}\theta _{o}} .
\end{eqnarray}
}
Then, in the equatorial plane $\theta _{o}=\pi /2$, the celestial coordinates
can be written as 
\begin{equation}
X^{2}+Y^{2}=\eta +\zeta ^{2}=R_{{\rm S}}^{2}.
\end{equation}
Here, $R_{{\rm S}}$ is the radius of the black hole shadow. We now use the equations (\ref{XX}) and (\ref{YY}) to plot the contour of the Euler-Heisenberg black hole shadow at the equatorial plane for different values of the quintessence parameter and dark matter in Fig.~\ref{photsf} and Fig.~\ref{photsfd}.     

Fig. \ref{photsf} shows that as the value of the dark matter parameter increases, the effective size of the shadow decreases. This decrease is larger for smaller quintessence state parameters. 

\begin{figure}[htb!]
\centering
\begin{minipage}[t]{0.43\textwidth}
        \centering
        \subcaption{{$\sigma= 0.1$}}
        \includegraphics[width=\textwidth]{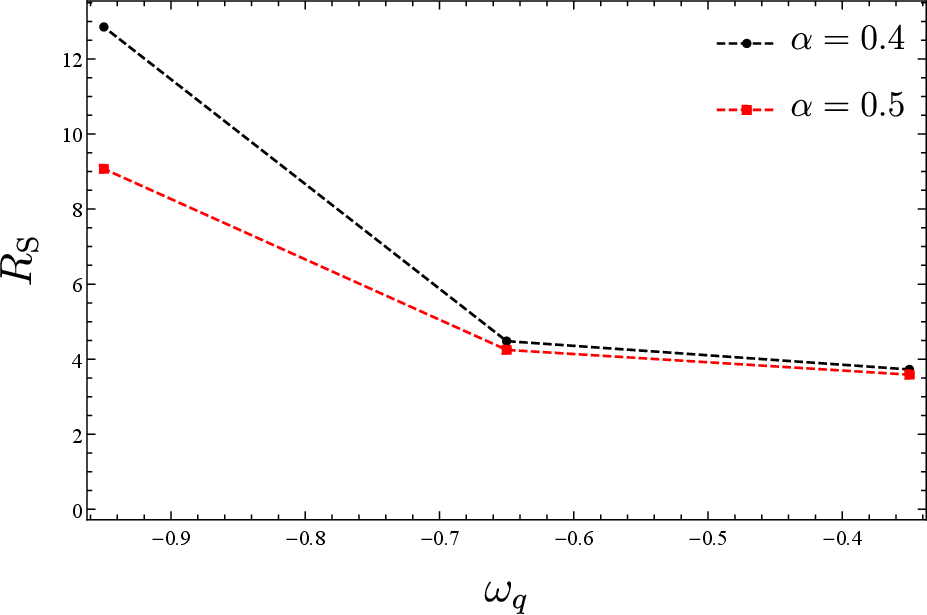}
         \label{fig:Ra}
\end{minipage}%
\begin{minipage}[t]{0.43\textwidth}
        \centering
        \subcaption{{$\alpha=0.2$}}
        \includegraphics[width=\textwidth]{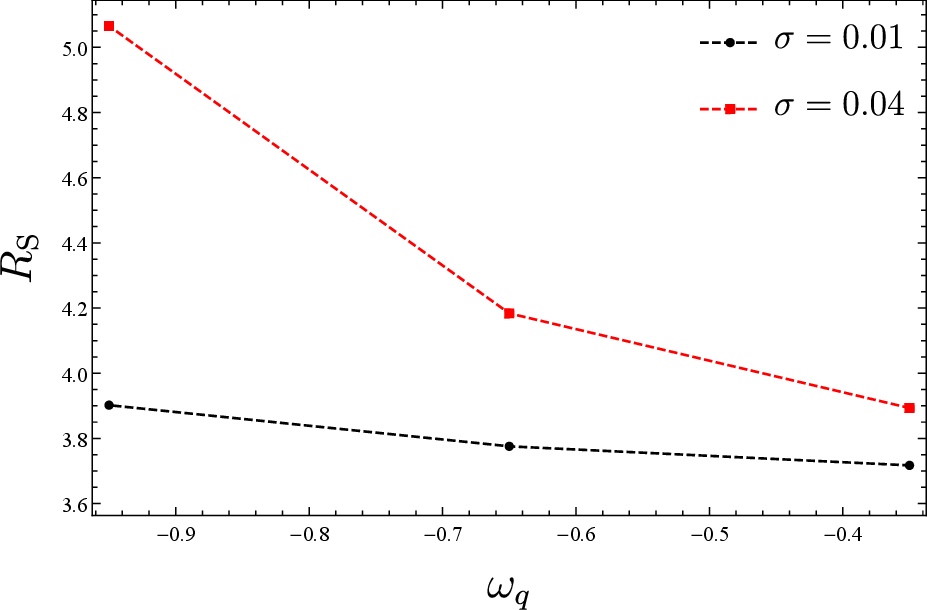}
       \label{fig:Rb}
   \end{minipage}%
\caption{{{\footnotesize Variation of the radius of the black hole shadow $R_{\rm S}$ versus $\omega_{q}$ for $M=1$, $q=0.5$, $a=0.1$. }}}
\label{radfg}
\end{figure}

Fig. \ref{photsfd} reveals the impact of the quintessence normalization factor. Unlike the dark matter intensity effect, the shadow radius increases as $\sigma$ increases. Moreover, we observe that this increase is greater at smaller $\omega_q$ values. To consolidate our results, we 
{ provide additional visualization. For example in Fig.~\ref{radfg}, we present a comparison of the changes in black hole shadow radius.  }

\begin{figure}[htb!]
\centering
\begin{minipage}[t]{0.43\textwidth}
        \centering
        \subcaption{{$\sigma= 0.1$}}
        \includegraphics[width=\textwidth]{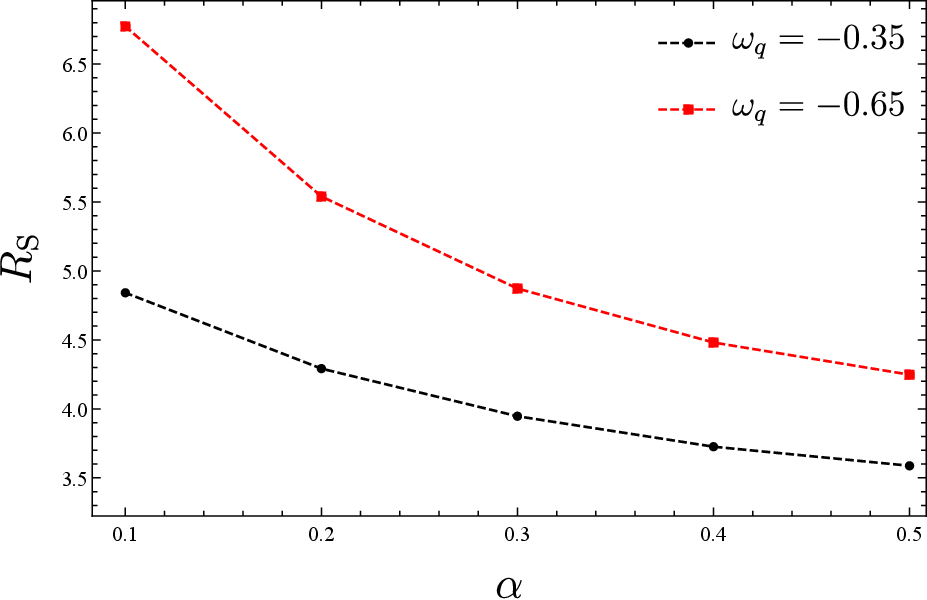}
               \label{fig:Raa}
\end{minipage}%
\begin{minipage}[t]{0.43\textwidth}
        \centering
        \subcaption{{$\alpha=0.2$}}
        \includegraphics[width=\textwidth]{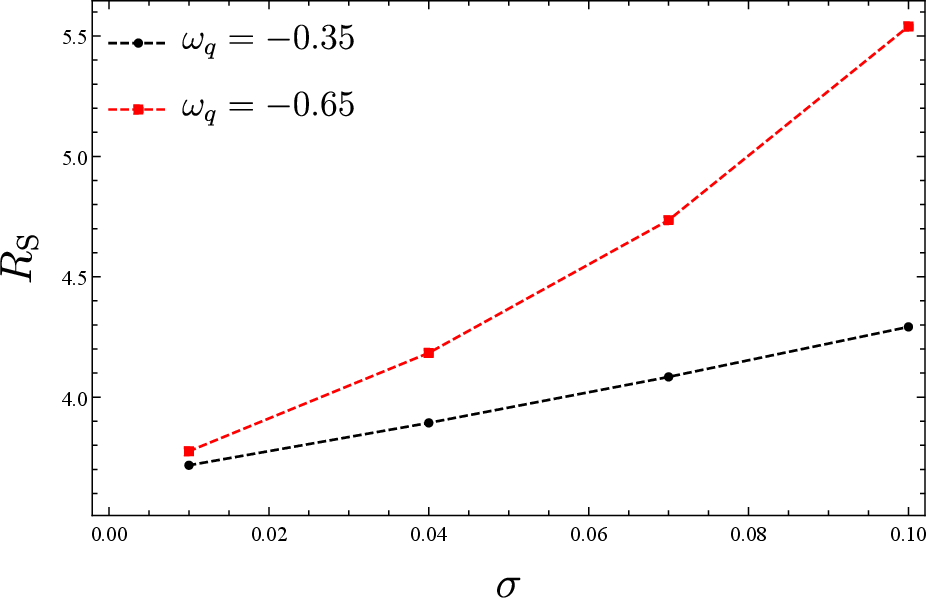}
       \label{fig:Rbb}
   \end{minipage}%
\caption{{{\footnotesize Variation of the radius of the black hole shadow $R_{\rm S}$ versus $\sigma$ /or $\alpha$ for $M=1$, $q=0.5$, $a=0.1$.}}}
\label{radfig}
\end{figure}

{ For a fixed value of sigma, we observe that as the quintessence state parameter increases the shadow radius decreases at a slower rate at larger alpha values. In contrast to this, for a fixed value of $\alpha$, we see that as $\omega_q$ increases, the shadow radius decreases at a greater rate at larger $\sigma$ values.  In Fig.~\ref{radfig}, we demonstrate a comparison of the alteration in shadow radius versus dark matter and energy parameters. }

{We observe that for a fixed value of $\sigma$, as $\alpha$ increases, the radius decreases at a greater rate for smaller values of the quintessence state parameter. In contrast, for a fixed value of $\alpha$, as $\sigma$ decreases, the radius increases at a greater rate for greater values of the quintessence state parameter. }

\newpage

\section{Quasinormal modes: Wentzel-Kramers-Brillouin  approximation} \label{sec:5}

\begin{figure}[htb!]
\centering
\begin{minipage}[t]{0.43\textwidth}
        \centering
        \subcaption{{$s=0$ and $\sigma=0.1$.}}
        \includegraphics[width=\textwidth]{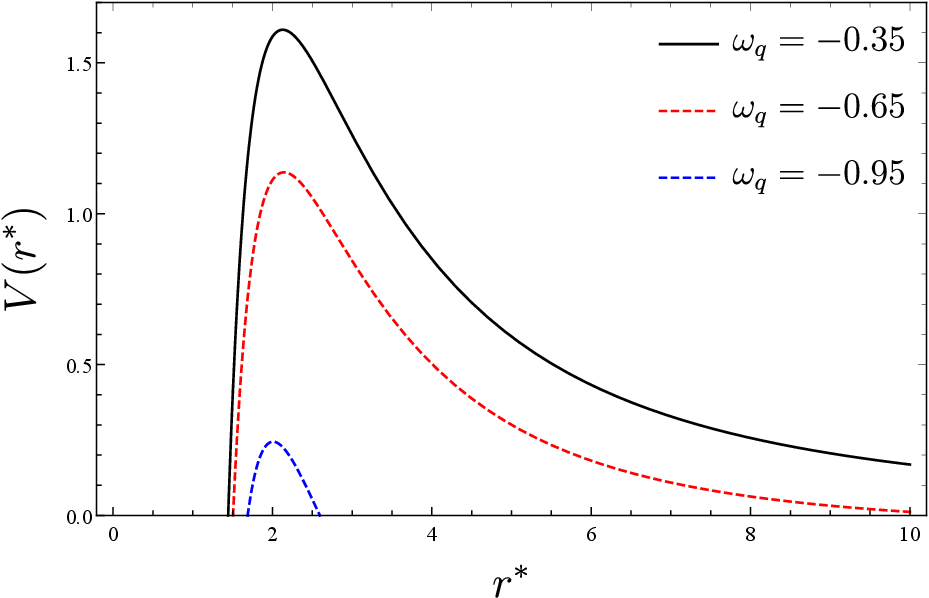}
         \label{fig:va}
\end{minipage}%
\begin{minipage}[t]{0.43\textwidth}
        \centering
        \subcaption{{$s=1$ and $\sigma=0.1$.}}
        \includegraphics[width=\textwidth]{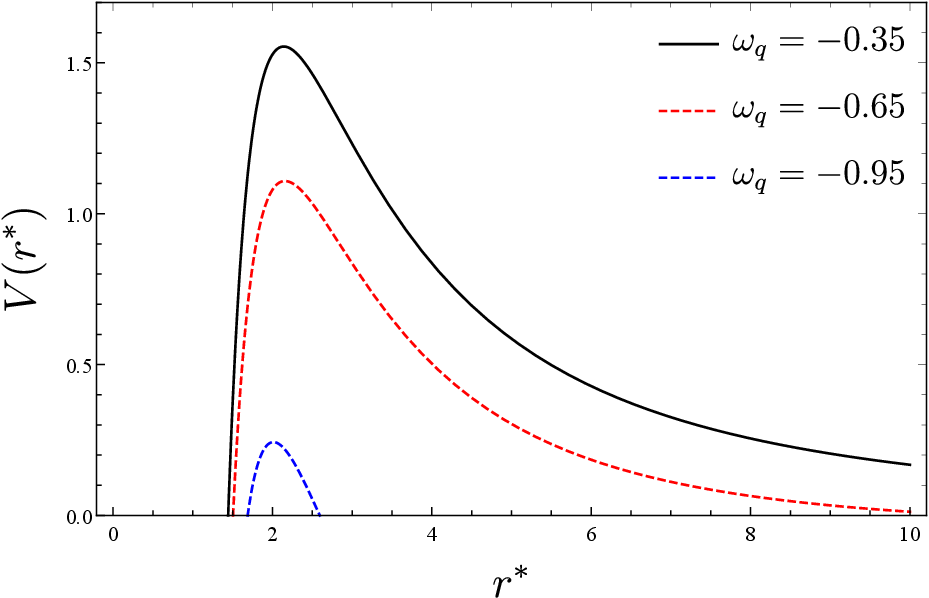}
       \label{fig:vb}
   \end{minipage}\\
\begin{minipage}[t]{0.43\textwidth}
        \centering
         \subcaption{{$s=0$ and $\omega_{q}=-0.35$.}}
        \includegraphics[width=\textwidth]{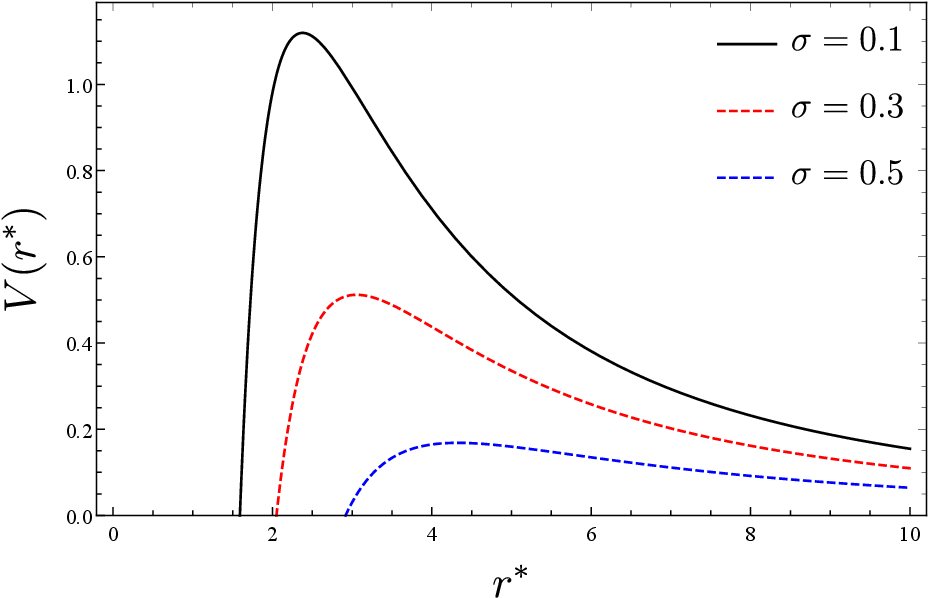}
        \label{fig:ve}
\end{minipage}%
\begin{minipage}[t]{0.43\textwidth}
        \centering
        \subcaption{{$s=1$ and $\omega_{q}=-0.35$.}}
        \includegraphics[width=\textwidth]{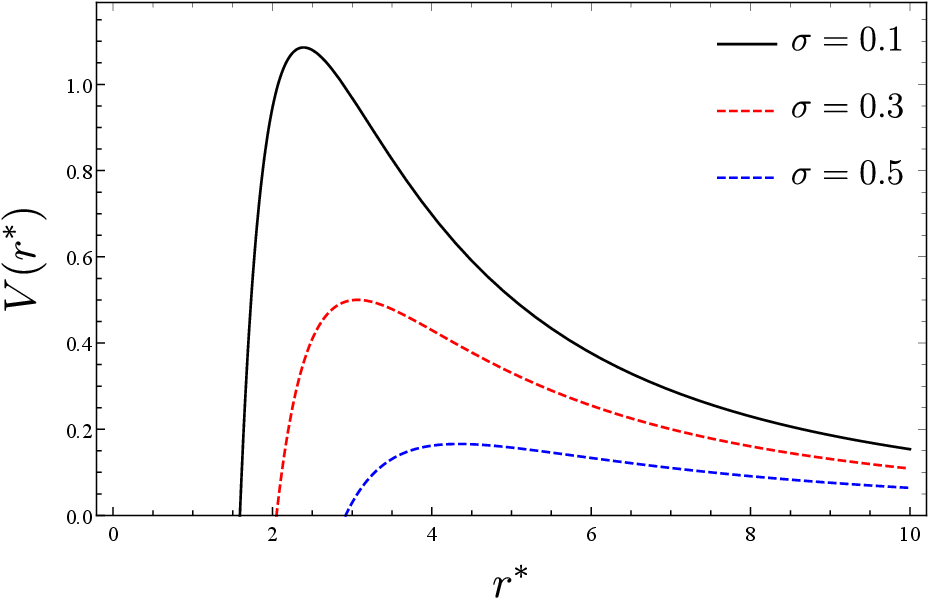}
       \label{fig:vf}
   \end{minipage}%

\caption{{{\footnotesize The impact of quintessence field on the potential for $M=1$, $\alpha=0.2$, $\ell=4$, $q=0.5$ and $a=0.1$.}}}
\label{veff1}
\end{figure}

In a curved spacetime, the behavior of a massless scalar field can be examined by the solutions of the Klein-Gordon equation, 
\begin{equation}
\frac{1}{\sqrt{-g}}\frac{\partial }{\partial x^{\mu }}\left( \sqrt{-g}g^{\mu
\nu }\frac{\partial }{\partial x^{\nu }}\psi \right) =0. \label{KGE}
\end{equation}%
For the 4-dimensional Euler-Heisenberg black hole spacetime Eq. (\ref{KGE}) becomes
\begin{equation}
-\frac{1}{f\left( r\right) }\frac{\partial ^{2}}{\partial t^{2}}\psi +\frac{1%
}{r^{2}}\frac{\partial }{\partial r}\left( r^{2}f\left( r\right) \frac{%
\partial }{\partial r}\psi \right) +\frac{1}{r^{2}}\left[ \frac{1}{\sin
\theta }\frac{\partial }{\partial \theta }\left( \sin \theta \frac{\partial 
}{\partial \theta }\right) +\frac{1}{\sin ^{2}\theta }\frac{\partial ^{2}}{%
\partial \varphi ^{2}}\right] \psi =0.  \label{qn}
\end{equation}
The field equation for vector and tensor perturbations obeys a field equation that is analogous to the one governing scalar frequencies. Due to the spherical symmetry of the Euler-Heisenberg black hole, the above equation can be solved by separating the variables. Consequently, the solution can be expressed as a decomposed function consisting of the product of separate functions that depend on the different coordinates $\left(t,r,\theta ,\varphi \right)$,
\begin{equation}
\psi \left( t,r,\theta ,\varphi \right) =\sum\limits_{\ell ,\mu }e^{-i\omega
t}\frac{\Psi _{\omega ,\ell }\left( r\right) }{r}\mathcal{Y}_{\ell ,\mu
}\left( \theta ,\varphi \right) ,  \label{dec}
\end{equation}%
where $\mathcal{Y}_{\ell ,\mu }\left( \theta ,\varphi \right) $ is the spherical harmonics and $\ell =0,1,2,...$ is the angular quantum number. Here, $\omega$ are the QNM frequencies, which are complex numbers that can be represented in the most general form of $\omega = \omega_{R} + i\omega_{I}$. In this representation, the real part, $\omega_R$, determines the normal frequency of the oscillations, while the imaginary part, $\omega_{I}$, represents the damping time, which is the rate at which the oscillations decay or diminish over time.
Using the tortoise coordinate,%
\begin{equation}
dr^{\ast }=\frac{dr}{f\left( r\right) },
\end{equation}
Eq.(\ref{qn}) reduces to the standard master equation of QNMs in static black hole spacetime%
\begin{equation}
\frac{d}{dr^{\ast }}\Psi _{\omega ,\ell }\left( r^{\ast }\right) -\left(
\omega ^{2}-V\left( r\right) \right) \Psi _{\omega ,\ell }\left( r^{\ast
}\right) =0,  \label{eqto}
\end{equation}%
with
\begin{equation}
V\left( r\right) =f\left( r\right) \left( \frac{\ell \left( \ell +1\right) }{%
r^{2}}+\frac{1-s^{2}}{r}\frac{df\left( r\right) }{dr}\right) .
\end{equation}
Here, $V\left( r\right)$
is the potential function, and $s = 0, 1$ corresponds respectively to scalar and electromagnetic perturbations, respectively. Before delving into the solutions of the field equation and determining the QNMs, let us first examine the potential related to perturbations for the $s=0,1$. To this end, we plot Fig.~\ref{veff1}  to demonstrate the variation of potential with 
the dark energy. 

\begin{figure}[htb!]
\centering
\begin{minipage}[t]{0.43\textwidth}
        \centering
        \subcaption{{$s=0$}}
        \includegraphics[width=\textwidth]{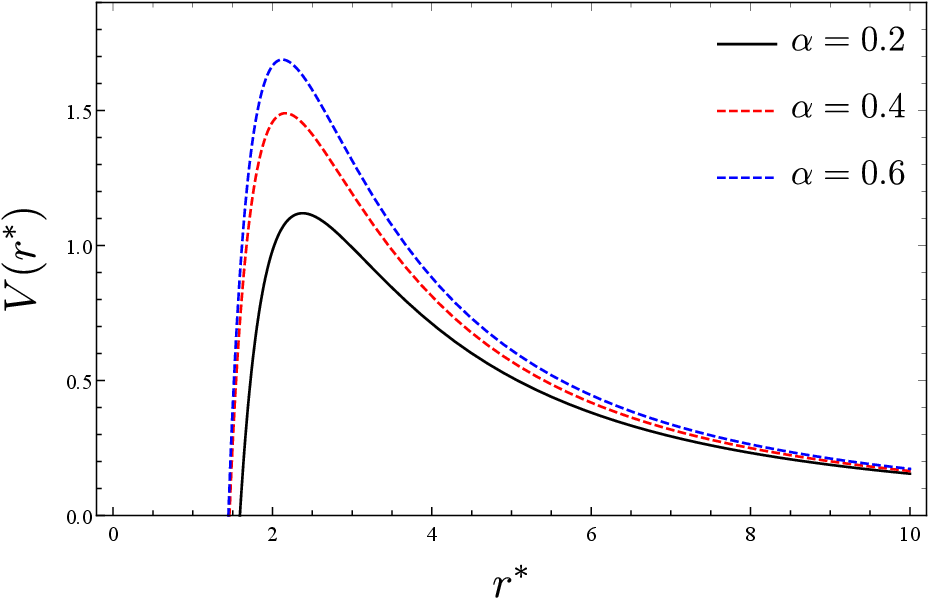}
         \label{fig:vc}
\end{minipage}%
\begin{minipage}[t]{0.43\textwidth}
        \centering
        \subcaption{{$s=1$}}
        \includegraphics[width=\textwidth]{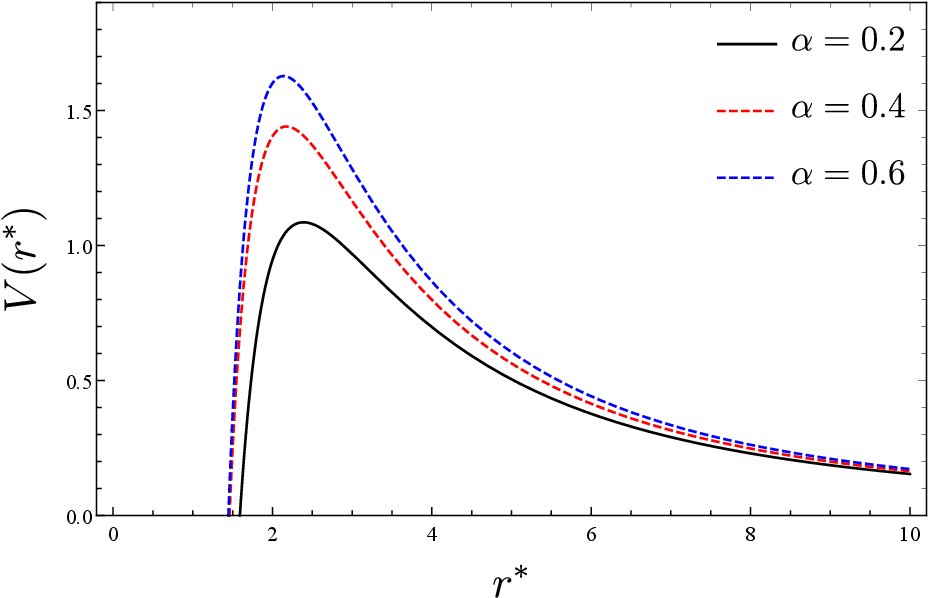}
       \label{fig:vd}
   \end{minipage}%
\caption{{{\footnotesize The impact of the dark matter field on the potential for  $M=1$, $\omega_{q}=-0.35$, $\sigma=0.1$, $\ell=4$,  $q=0.5$ and $a=0.1$.}}}
\label{veff2}
\end{figure}

Fig.~\ref{veff1} reveals that the peak of the Euler-Heisenberg black hole potential essentially differs for various values of $\omega_{q}$ and $\sigma$. For example, as $\omega_{q}$ decreases the peak value of the potential decreases.  As the quintessence normalization parameter increases, the potential damps at a larger interval. This characteristic behavior almost mimics itself for each value of $s$. Then, we examine the effect of dark matter intensity in Fig.~\ref{veff2}.

\newpage

In Fig.~\ref{veff2}, we observe that for larger values of dark matter intensity the height of the potential peak increases. We note that this impact is in contrast with the quintessence field effect.

Now, we focus on finding a solution to Eq.~(\ref{eqto}) with appropriate boundary conditions. In this particular case, the acceptable solutions are those that represent purely
ingoing waves near the event horizon of the black hole,
\begin{equation}
\Psi \sim \left\{ 
\begin{array}{c}
e^{-i\omega r^{\ast }};r^{\ast }\rightarrow -\infty  \\ 
e^{i\omega r^{\ast }};r^{\ast }\rightarrow +\infty 
\end{array}%
\right. ,
\end{equation}%
that means the ingoing wave at the horizon and the outgoing wave at infinity. 

Over the past few decades, the QNMs have been investigated using various methods \cite{Alex11}. Each method has its strengths and weaknesses. Developing more efficient techniques to study QNMs is an active area of ongoing research. In the present study, we employ the semi-analytical Wentzel-Kramers-Brillouin (WKB) approximation method. This technique involves matching the WKB expansion of the modes at the event horizon and spatial infinity with the Taylor expansion of the effective potential near the peak of the potential barrier. The WKB method was initially proposed in \cite{Schutz, Mashhoon, Blome, Liu} and subsequently extended to higher-order approximations,
including the 6th order \cite{Iyer, Konoplya, Konoplyar} and 13th order \cite{Matyjasek, Konoplya13}. It is noteworthy that
increasing the order of the WKB approximation does not always result in a better approximation for the frequency. Consequently, we have chosen to utilize the 6th order expansion for our study, which is represented by the following equation,
\begin{equation}
i\frac{\left( \omega_{n} -V_{0}\right) }{\sqrt{-2V_{0}"}}-\sum_{i=2}^{N}\Lambda
_{i}=n+\frac{1}{2}.  \label{qnmf}
\end{equation}
Here, $V_{0}$ represents the maximum height of the potential, while $V_{0}"$ denotes the second derivative of the potential with respect to the tortoise coordinate. $\Lambda _{i}$ is a constant coefficient arising from higher-order WKB approximation method corrections, and the variable $n=0,1,2,...$ is the overtone number. The explicit expressions for the higher-order $\Lambda _{i}$ coefficients are provided in \cite{Konoplya}. Since Eq.~(\ref{qnmf}) involves non-trivial functions of physical parameters, we have computed the QNM frequencies for different values of these parameters. Here, we consider the massless scalar field functions and electromagnetic (massless vector) field which are subject to the condition $\ell>n$. In Table~\ref{tab:tests0} and Table~\ref{tabs1}, we list the QNMs frequencies for different values of $\omega_{q}$.  The first observation is that the imaginary parts of the QNM frequencies are negative across all branches or modes. This behavior characterizes the stable propagation of scalar fields in this background.
\newpage
\begin{table}[tbph]
\centering%
\caption{{{\footnotesize The QNMs of the Euler-Heisenberg black hole for the massless scalar perturbation with $M = 1$, $\sigma= 0.1$, $\alpha = 0.2$, $a = 0.1$ and $q=0.5$  via the 6th order WKB approximation method.}}}
\begin{tabular}{l|l|l|l|l}
\hline
\rowcolor{lightgray} $\ell $ & $n$ & $\omega _{q}=-0.35$ & $\omega _{q}=-0.65
$ & $\omega _{q}=-0.95$ \\ \hline
3 & 0 & 0.821849 - 0.119525 i & 0.631236 - 0.085595 i & 0.059110 - 0.409985 i  \\ 
& 1 & 0.834976 - 0.386800 i & 0.640894 - 0.272814 i & 0.185721 - 0.420349 i  \\ 
& 2 & 0.917164 - 0.591183 i  & 0.692288 - 0.427960 i & 0.292899 - 0.447585 i \\ \hline
4 & 0 & 1.052390 - 0.117290 i & 0.811230 - 0.083990 i & 0.058759 - 0.531230 i \\ 
& 1 & 1.061280 - 0.367746 i & 0.816844 - 0.261521 i & 0.181935 - 0.538541 i \\ 
& 2 & 1.107400 - 0.669605 i & 0.847318 - 0.465074 i & 0.297566 - 0.556725 i \\ 
& 3 & 1.195490 - 0.816909 i & 0.902850 - 0.591916 i & 0.410659 - 0.589720 i \\ \hline
5& 0 & 1.284140 - 0.115956 i & 0.991613 - 0.083064 i & 0.058567 - 0.652094 i\\
& 1 & 1.289160 - 0.359118 i & 0.994541 - 0.256160 i & 0.179024 - 0.657312 i \\
& 2 & 1.319690 - 0.627952 i & 1.013550 - 0.443819 i & 0.298187 - 0.669799 i \\
& 3 & 1.389320 - 0.831201 i & 1.057550 - 0.599333 i & 0.416561 - 0.695291 i \\
& 4 & 1.475950 - 1.041820 i & 1.114530 - 0.755441 i & 0.527534 - 0.731896 i\\
\hline\hline
\end{tabular}%
\label{tab:tests0}
\end{table}
\begin{table}[tbph]
\centering%
\caption{{{\footnotesize The QNMs of the Euler-Heisenberg black hole for the electromagnetic (massless vector) perturbation with $M = 1$, $\sigma= 0.1$, $\alpha = 0.2$, $a = 0.1$ and $q=0.5$  via the 6th order WKB approximation method.}}}
\begin{tabular}{l|l|l|l|l}
\hline
\rowcolor{lightgray} $\ell $ & $n$ & $\omega _{q}=-0.35$ & $\omega _{q}=-0.65
$ & $\omega _{q}=-0.95$ \\ \hline
3 & 0 & 0.796577 - 0.116386 i & 0.618480 - 0.083180 i & 0.058183 - 0.410959 i \\ 
& 1 & 0.811524 - 0.373921 i & 0.616869 - 0.275347 i & 0.174668 - 0.413403 i \\ 
& 2 & 0.655372 - 0.691307 i & 0.536378 - 0.509286 i & 0.291486 - 0.420788 i \\ \hline
4 & 0 & 1.033650 - 0.114800 i & 0.801863 - 0.082202 i & 0.058119 - 0.532451 i \\ 
& 1 & 1.032210 - 0.362272 i & 0.797656 - 0.258540 i & 0.174457 - 0.534227 i \\ 
& 2 & 0.995891 - 0.667079 i & 0.784227 - 0.474995 i & 0.291028 - 0.537822 i  \\ 
& 3 & 0.783103 - 0.965639 i & 0.655831 - 0.718252 i & 0.408217 - 0.542989 i  \\ \hline
5 & 0 & 1.269290 - 0.114005 i & 0.984240 - 0.081709 i & 0.058087 - 0.653273 i\\ 
& 1 & 1.263400 - 0.353612 i & 0.979496 - 0.252089 i & 0.174293 - 0.654604 i\\ 
& 2 & 1.249800 - 0.630603 i & 0.974352 - 0.447751 i & 0.290717 - 0.657759 i \\ 
& 3 & 1.178610 - 0.969877 i & 0.940628 - 0.689618 i & 0.407475 - 0.662160 i \\ 
& 4 & 0.899207 - 1.222810 i & 0.762905 - 0.920460 i & 0.524736 - 0.667941 i \\ 
\hline\hline
\end{tabular}%
\label{tabs1}
\end{table}

\begin{figure}[htb!]
\centering
\begin{minipage}[t]{0.43\textwidth}
        \centering
        \includegraphics[width=\textwidth]{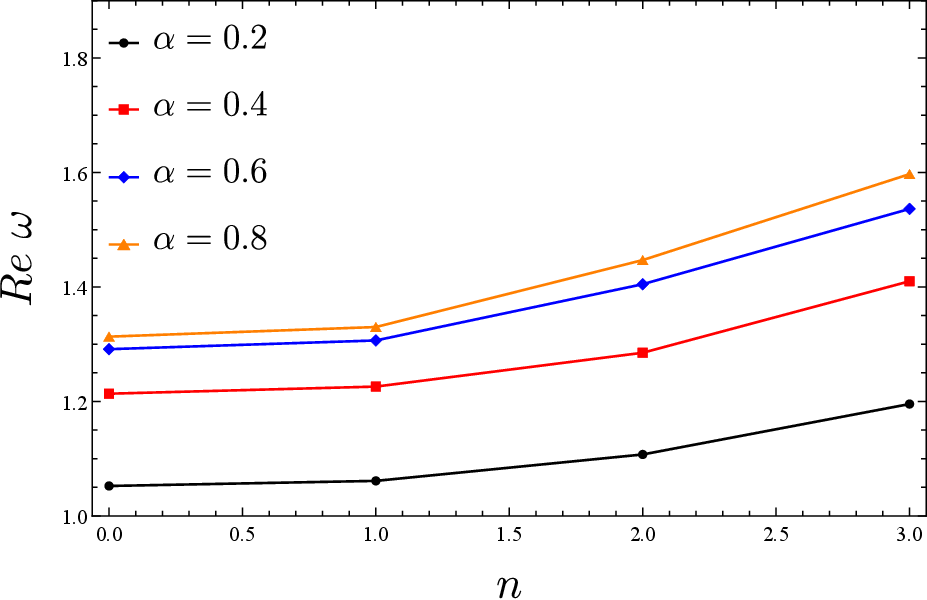}
\end{minipage}%
\begin{minipage}[t]{0.43\textwidth}
        \centering        \includegraphics[width=\textwidth]{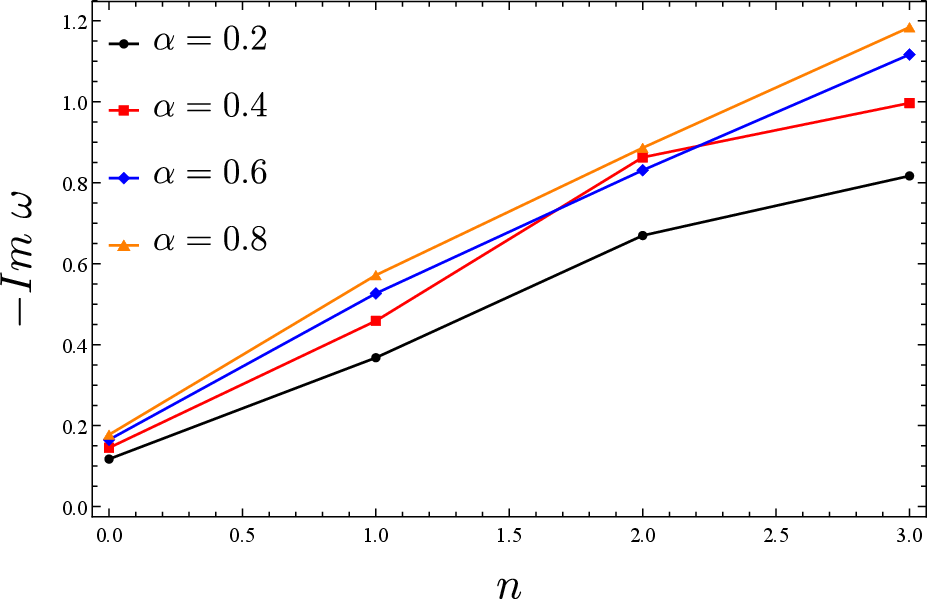}
   \end{minipage}%
\caption{{{\footnotesize The variation of the Re $\omega$ and Im $\omega$  for the massless scalar field versus overtone numbers for different values of dark matter with $M=1$, $q=0.5$, $\omega_{q}=-0.35$, $\sigma=0.1$, $\ell=4$ and $a=0.1$.}}}
\label{fig:q1}
\end{figure}

\begin{figure}[htb!]
\centering
\begin{minipage}[t]{0.43\textwidth}
        \centering
        \includegraphics[width=\textwidth]{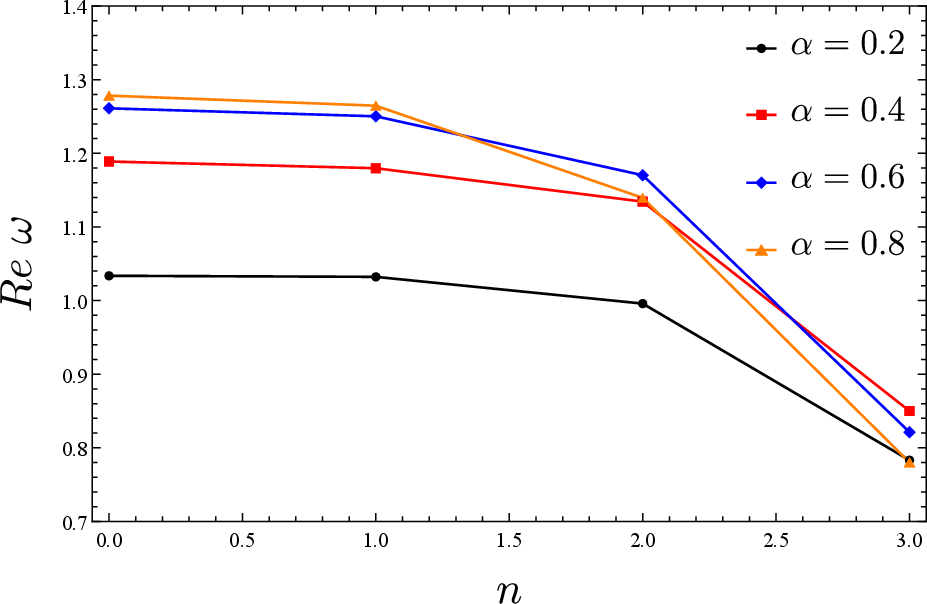}
        \end{minipage}%
\begin{minipage}[t]{0.43\textwidth}
        \centering
        \includegraphics[width=\textwidth]{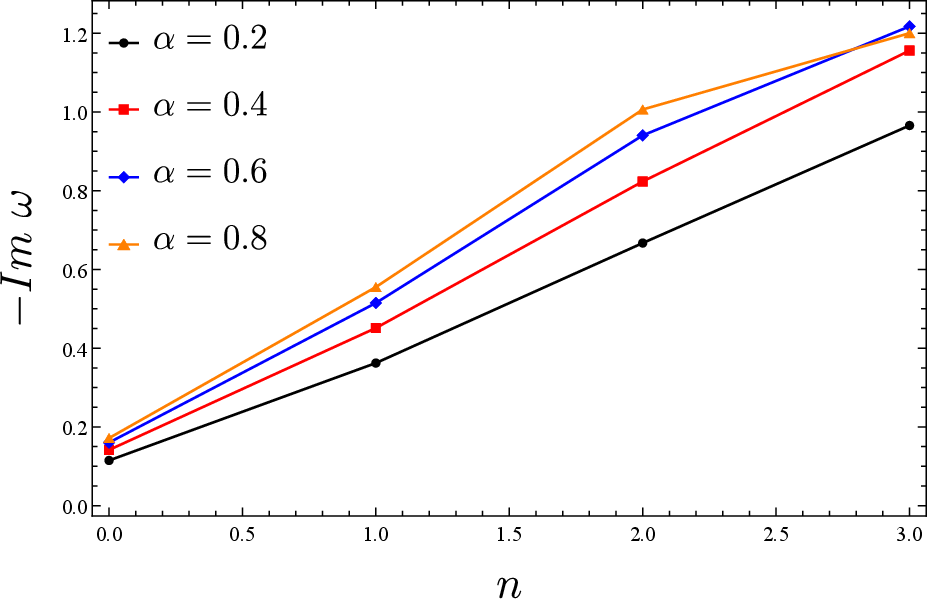}
          \end{minipage}%
\caption{ {{\footnotesize The variation of the Re $\omega$ and Im $\omega$  for the electromagnetic (massless vector) perturbation versus overtone numbers for different values of dark matter with $M=1$, $q=0.5$, $\omega_{q}=-0.35$, $\sigma=0.1$ and $a=0.1$.}}}
\label{fig:q2}
\end{figure}

\begin{figure}[htb!]
\centering
\begin{minipage}[t]{0.43\textwidth}
        \centering
        \includegraphics[width=\textwidth]{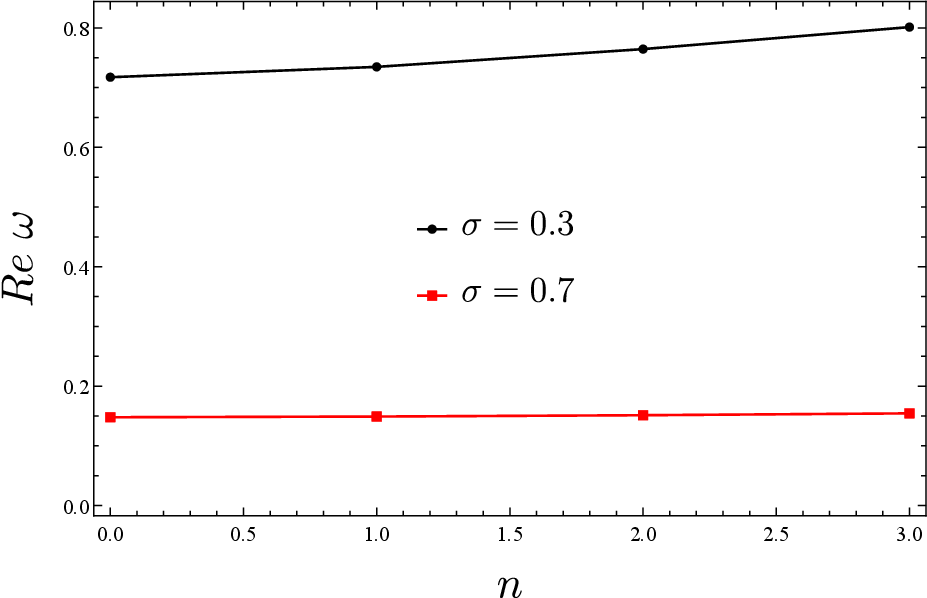}
\end{minipage}%
\begin{minipage}[t]{0.43\textwidth}
        \centering
        \includegraphics[width=\textwidth]{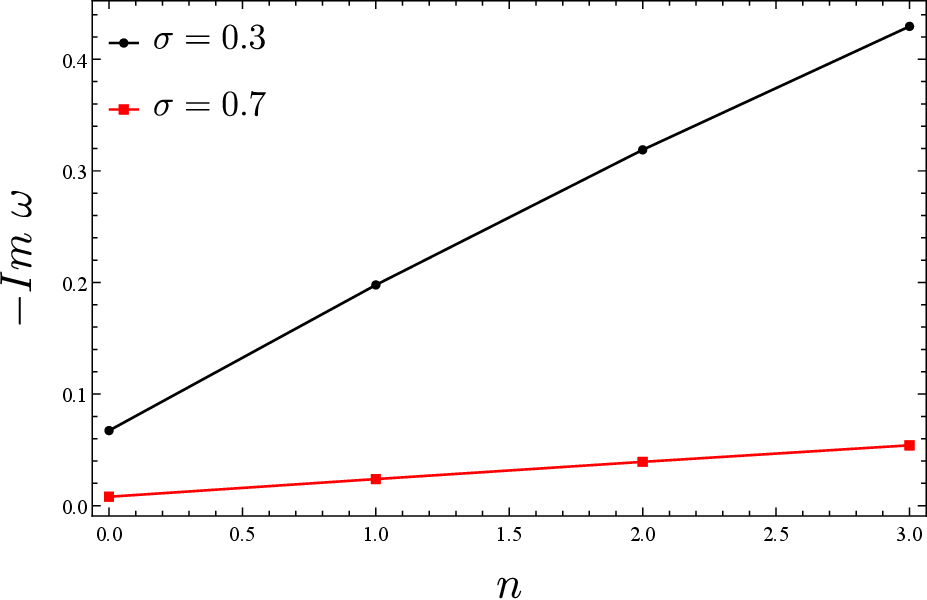}
       \label{fig:q5}
   \end{minipage}%
\caption{ {{\footnotesize The variation of the Re $\omega$ and Im $\omega$  for the massless scalar perturbation versus overtone numbers for different values of normalization factor with $M=1$, $q=0.5$, $\omega_{q}=-0.35$,$\alpha=0.2$ and $a=0.1$.}}}
\label{fig:q3}
\end{figure}

\begin{figure}[htb!]
\centering
\begin{minipage}[t]{0.43\textwidth}
        \centering
        \includegraphics[width=\textwidth]{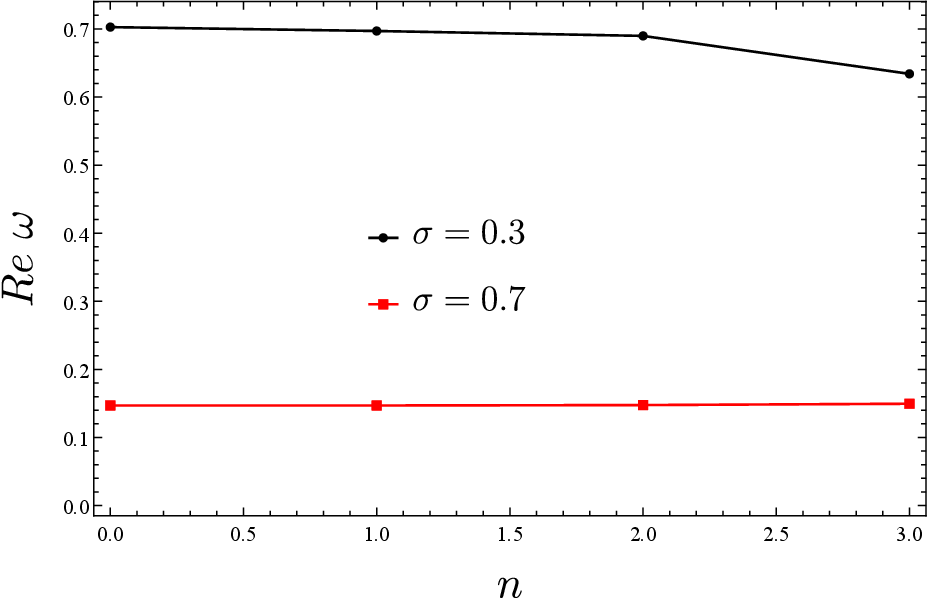}
\end{minipage}%
\begin{minipage}[t]{0.43\textwidth}
        \centering
        \includegraphics[width=\textwidth]{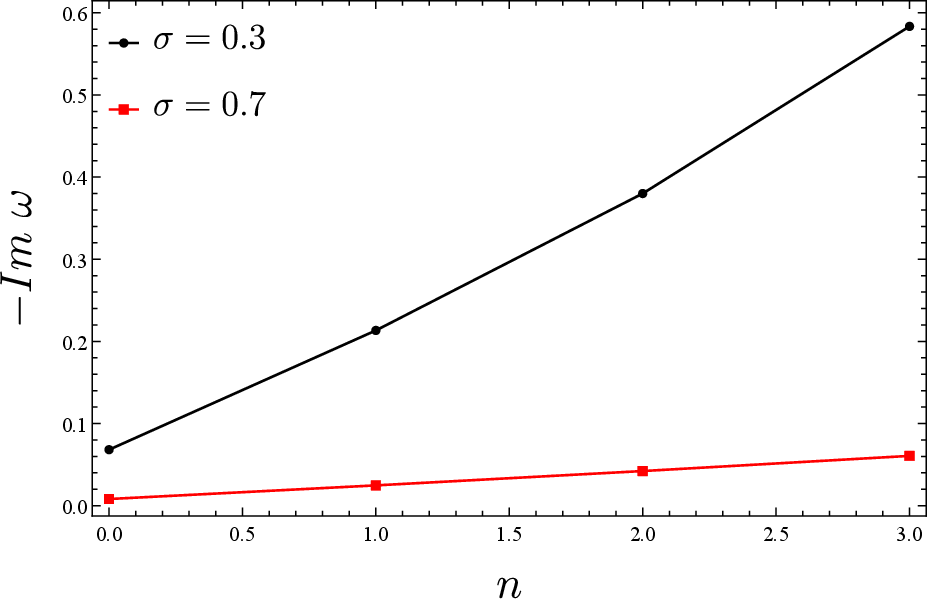}
   \end{minipage}%
\caption{{{\footnotesize The variation of the Re $\omega$ and Im $\omega$  for the electromagnetic (massless vector) perturbation versus overtone numbers for different values of normalization factor with $M=1$, $q=0.5$, $\omega_{q}=-0.35$, $\alpha=0.2$ and $a=0.1$.}}}
\label{fig:q6}
\end{figure}

\begin{figure}[htb!]
\centering
\begin{minipage}[t]{0.43\textwidth}
        \centering
        \includegraphics[width=\textwidth]{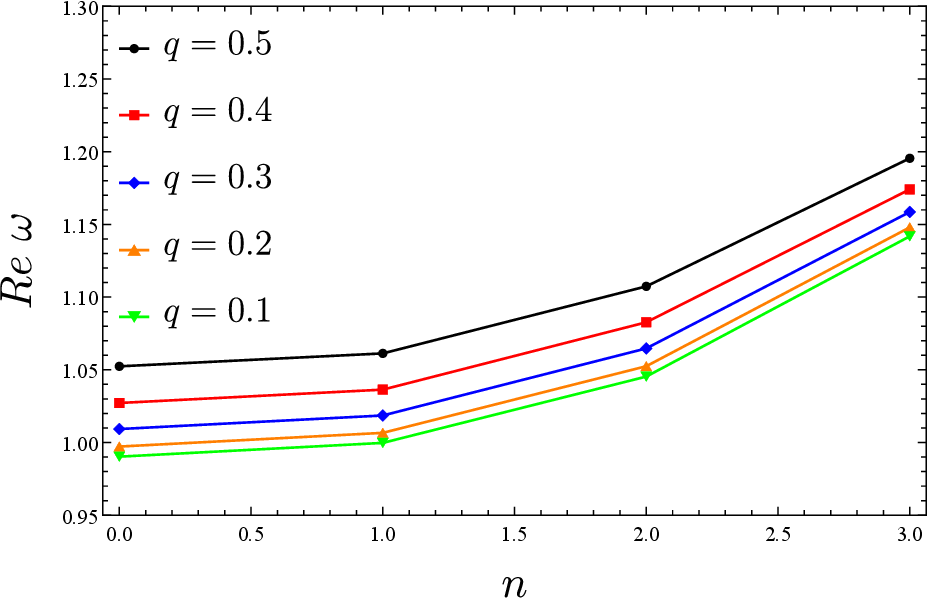}
\end{minipage}%
\begin{minipage}[t]{0.43\textwidth}
        \centering
        \includegraphics[width=\textwidth]{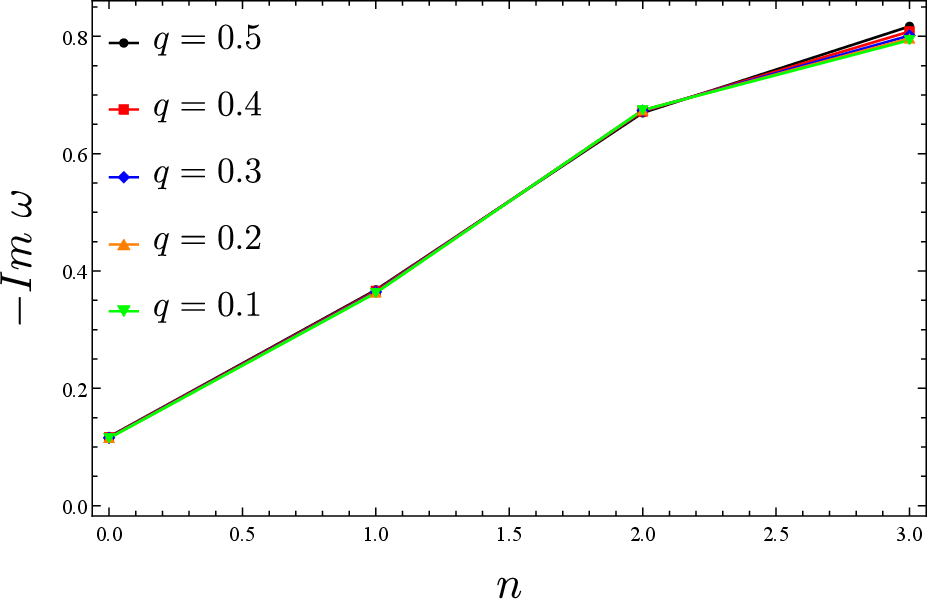}
   \end{minipage}%
\caption{{{\footnotesize The variation of the Re $\omega$ and Im $\omega$  for the massless scalar perturbation versus overtone numbers for different values of $q$ with $M=1$, $\sigma=0.1$, $\omega_{q}=-0.35$, $\alpha=0.2$ and $a=0.1$.}}}
\label{fig:q7}
\end{figure}

\begin{figure}[htb!]
\centering
\begin{minipage}[t]{0.43\textwidth}
        \centering
        \includegraphics[width=\textwidth]{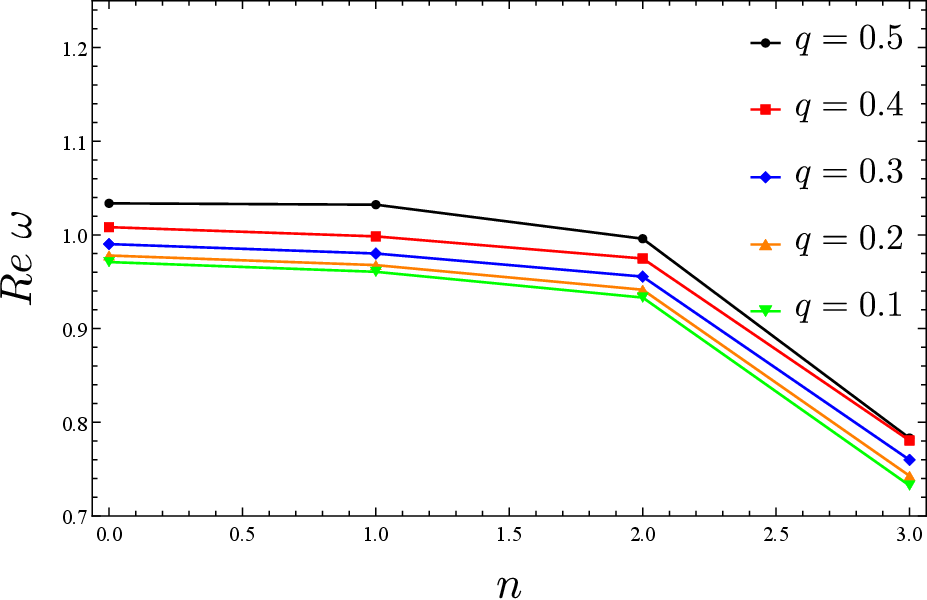}
\end{minipage}%
\begin{minipage}[t]{0.43\textwidth}
        \centering
        \includegraphics[width=\textwidth]{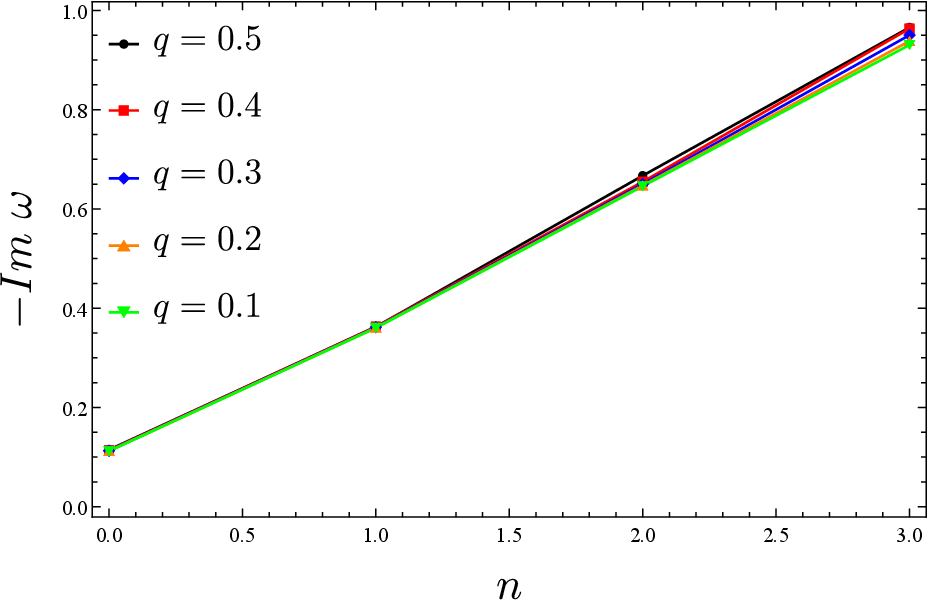}
   \end{minipage}%
\caption{{{\footnotesize The variation of the Re $\omega$ and Im $\omega$  for the electromagnetic (massless vector) perturbation versus overtone numbers for different values of $q$ with $M=1$, $\sigma=0.1$, $\omega_{q}=-0.35$, $\alpha=0.2$ and $a=0.1$.}}}
\label{fig:q8}
\end{figure}

Finally, we analyze our findings with graphical demonstrations in between Fig.~\ref{fig:q1} -  Fig.~\ref{fig:q8}.

\newpage

\noindent Fig.~\ref{fig:q1} illustrates how dark matter changes the real part {(Re$\omega$)} and the negative imaginary part {(-Im$\omega$)} of the frequency. We observe that the real part becomes larger as the value of $n$ increases. We note that the plots for the negative imaginary part show a similar trend. 

\noindent Fig.~\ref{fig:q2} shows the behavior for electromagnetic (massless vector) perturbations. In this case, the real part decreases as $n$ increases, which contrasts with the scalar field. However, the graphs for the negative imaginary part display the same patterns between the scalar and vector cases.

\noindent Fig.~\ref{fig:q3} reveals the impact of the quintessence normalization factor on the real and imaginary parts of QNMs in the massless scalar field case.  Here, we observe that both parts increase as the overtone number increases. This increase is relatively greater at small $\sigma$ values. 

\noindent Fig.~\ref{fig:q6} shows that in the electromagnetic perturbation case, the real sector decreases while the imaginary sector increases as the overtone number increases. Here, we note that this impact becomes negligible at relatively large quintessence normalization parameters. 

Fig.~\ref{fig:q7} and Fig.~\ref{fig:q8} show the effect of $q$ for the scalar and electromagnetic cases respectively. In particular, 
as the overtone numbers increase, the real sector increases in the scalar case, while it decreases in the electromagnetic case. These changes are of greater magnitude as the parameter $q$ assumes larger values. On the other hand, the imaginary sector increases monotonically in both cases, and a change in the $q$ parameter only affects its values at relatively large overtone numbers.  


\section{Conclusions}
In this manuscript, we have investigated the Euler-Heisenberg black hole in the presence of the quintessence matter field and the perfect fluid dark matter. To this end, we first derived the lapse function and then discussed the effect of the dark sector components in three different scenarios in the rest of the manuscript. The impact of quintessence matter on the mass function at the horizon showed that the black hole mass does not increase monotonically in all scenarios. The greater quintessence matter imposes an upper limit on the horizon values of the black hole, so it changes the black hole characteristics. However, we found that dark matter does not have such a change in black hole characteristics and only alters the physically meaningful range of the event horizon. Then, we derived the Hawking temperature. We found that the Hawking temperature first increases and then decreases, and the quintessence normalization factor plays an important role in this behavior. We also concluded that dark matter intensity has a crucial effect on the temperature only at relatively small horizons. We then showed that the horizon entropy does not depend on either the quintessence or the dark matter. Our detailed analysis of the specific heat functions showed that the black hole can have unstable, stable, and unstable phases for some values of the quintessence matter field. For those cases, we calculated the remnant mass values. We then found that with sufficiently large dark matter intensities, the black hole has no phase transition. After that, we obtained the equation of state and plotted the pressure isotherms to discuss their characteristics in different scenarios. We also found that dark matter affects the pressure isotherms at relatively small horizons. We finished the investigation on the thermal quantities by tabulating the critical values of horizon, temperature, and pressure.  We then focused on the black hole shadow. After deriving the effective potential we calculated the photon and shadow radii. We concluded that as the dark matter intensity increases the effective shadow radius decreases.  We showed that the quintessence matter field impact has a different impact. For example, an increase in the quintessence normalization parameter leads to an increase in the shadow radius. Finally, we studied the quasinormal modes including the 6th order in  WKB approximation. After tabulating the modes, we graphically showed the impact of quintessence and dark matter fields on these modes. In conclusion, we found that the simultaneous presence of dark energy and dark matter leads to very interesting effects on the thermodynamics, shadows, and quasinormal modes of Euler-Heisenberg black holes.

\section*{Conflict of Interests} 

The authors declare no such conflict of interest.

\section*{Data Availability Statement} 

No data were generated or analyzed in this study.

\section*{Acknowledgements}
{The authors are thankful to the anonymous referee for the constructive comments.} B. C. L. is grateful to Excellence Project PřF UHK 2211/2023-2024 for the financial support.


\begin{thebibliography}{}
\bibitem{Einstein} A. Einstein, Annalen der Physik \textbf{49}, 769 (1916).
\bibitem{Bekenstein} J. D. Bekenstein, Phys. Rev. D \textbf{7}, 2333 (1973).
\bibitem{Hawking} S. W. Hawking, Commun. Math. Phys. \textbf{43}, 199 (1975).
\bibitem{Hawking1} S. W. Hawking,Phys. Rev. D \textbf{13}, 191 (1976).
\bibitem{Bardeen} J. M. Bardeen, B. Carter, and S. W. Hawking, Commun. Math. Phys. \textbf{31}, 161 (1973).
\bibitem{Riess}  A. G. Riess et al., Astron. J. \textbf{116}, 1009 (1998).
\bibitem{Ries}A. G. Riess et al.,  Astrophys. J. \textbf{607}, 665 (2004).
\bibitem{Rss}S. Perlmutter et al., Astrophys. J. \textbf{517}, 565 (1999).
\bibitem{Knop}R.  A . Knop et al., Astrophys. J. \textbf{598}, 102 (2003).
\bibitem{Miller} A. D. Miller et al., Astrophys. J. Lett. \textbf{524}, L1 (1999)
\bibitem{Hanany} S. Hanany et al., Astrophys. J. Lett. \textbf{545}, L5 (2000).
\bibitem{Mason}B. S. Mason et al., Astrophys. J. \textbf{591}, 540 (2003).    
\bibitem{Page} L. Page L et al.,  Astrophys. J. Suppl. \textbf{148}, 233 (2003).
\bibitem{Scranton} R. Scranton et al., arXiv:astro-ph/0307335.
\bibitem{Tegmark}M. Tegmark et al., Phys. Rev. D \textbf{69}, 103501 (2004).
\bibitem{Peebles} P. J. E. Peebles  and B. Ratra, Rev. Mod. Phys. \textbf{75}, 559 (2003).
\bibitem{Padmanabhan} T. Padmanabhan, Phys. Rep. \textbf{380}, 235 (2003).
\bibitem{Padmanabhan1}T. Padmanabhan, Curr. Sci. \textbf{88}, 1057 (2005).
\bibitem{Weinberg}S. Weinberg, Rev. Mod. Phys. \textbf{61}, 1 (1989).
\bibitem{Carroll}S. M. Carroll, Phys. Rev. Lett. \textbf{81}, 3067 (1998).





\bibitem{Rubin1970} V. C. Rubin, W. K. Ford, Astrophys. J. \textbf{159}, 379 (1970). 

\bibitem{Roberts1973} M. S. Roberts, A. H. Rots, Astron. Astrophys.  \textbf{26}, 483 (1973).

\bibitem{Rubin1980} V. C. Rubin, W. K. Ford Jr., N. Thonnard, Astrophys. J. \textbf{238}, 471 (1980).

\bibitem{Ostriker1973} J. P. Ostriker, P. J. E. Peebles, Astrophys. J.
\textbf{186}, 467 (1973).

\bibitem{Einasto1974} J. Einasto, A Kaasik, F. Saar, Nature \textbf{250}, 309 (1974).

\bibitem{Davis1985} M. Davis, G. Efstathiou, C. S. Frenk,  S. D. M. White,  Astrophys. J. \textbf{292}, 371 (1985).

\bibitem{Clowe2006} D. Clowe, M. Bradac, A. H. Gonzalez, M. Markevitch, S. W. Randall, C. Jones, D. Zaritsky,  Astrophys. J. \textbf{648}, L109 (2006).

\bibitem{Mandel2013} R. Mandelbaum  et al., Mon. Not. R. Astron. Soc. \textbf{432}, 1544 (2013).

\bibitem{Hinshaw2013} G. Hinshaw et al.,  (WMAP Collaboration)  Astrophys. J. Suppl.  \textbf{208}, 19 (2013).

\bibitem{Ade2014} P. A. R. Ade et al., (Planck Collaboration), Astron. Astrophys. \textbf{571}, A16 (2014).

\bibitem{Komatsu2011} E. Komatsu et al., (WMAP Collaboration), Astrophys. J. Suppl. \textbf{192}, 18 (2011).



\bibitem{Bertone2018} G. Bertone, D. Hooper, Rev. Mod. Phys. \textbf{90}, 045002 (2018).

\bibitem{Huterer2018} D. Huterer, D. L. Shafer, Rep. Prog. Phys. \textbf{81}, 016901 (2018).

\bibitem{Navarro1996} J. F. Navarro, C. S. Frenk, S. D. M. White, Astrophys. J. \textbf{462}, 563 (1996). 


\bibitem{Boehm2004} C. Boehm, P. Fayet, Nucl. Phys. B \textbf{683}, 219 (2004).

\bibitem{Bertone2005} G. Bertone, D. Hooper, J. Silk,
Phys. Rept. \textbf{405}, 279 (2005). 

\bibitem{Feng2009} J. L. Feng, M. Kaplinghat, H. Tu, H. -B. Yu,  JCAP \textbf{2009},
004 (2009). 


\bibitem{Graham2015} P. W. Graham, I. G. Irastorza, S. K. Lamoreaux, A. Lindner,  K. A. van Bibber, Ann. Rev. Nucl. Part. Sci. \textbf{65}, 485 (2015).

\bibitem{Schumann2019} M. Schumann,  J. Phys. G:
Nucl. Part. Phys. \textbf{46}, 103003 (2019). 

\bibitem{Boyarsky2019} A. Boyarsky, M. Drewes, T. Lasserre, S. Mertens, O. Ruchayskiy, Prog. Part. Nucl. Phys. \textbf{104}, 1 (2019). 


\bibitem{Deur2019} A. Deur, Eur. Phys. J. C \textbf{79}, 883 (2019).
{
\bibitem{Yaholom2020}  A. Yahalom, Symmet. \textbf{12}, 1693 (2020).}

\bibitem{Oks2020} E. Oks,  Res. Astron. Astrophys. \textbf{20}, 109 (2020).

\bibitem{Qiao2021} C. -K. Qiao, S. -T. Lin, H. -C. Chi, H. -T. Jia, J. High Energ. Phys. \textbf{2021}, 184 (2021). 

\bibitem{Oks2024} E. Oks,  New Astron. \textbf{107}, 102134 (2024).

\bibitem{Ruiz2021} J. A. Ruiz, Eur. Phys. J Plus \textbf{136}, 33 (2021).

{
\bibitem{Abbas2023} G. Abbas,  R. H. Ali, Eur. Phys. J. C \textbf{83}, 407 (2023).

\bibitem{Chen2024} H. Chen, S. H. Dong, S. Hassanabadi, N. Heidari, H. Hassanabadi, Chinese Phys. C \textbf{48}, 085105(2024).

\bibitem{Kiselev2002} V. V. Kiselev, arXiv:gr-qc/0303031.

\bibitem{Kiselev2003} V. V. Kiselev, Class. Quantum Grav. \textbf{20}, 1187 (2003).
}



\bibitem{Sidd2003} G. Siddharta, T. Matos, D. Nunez, E. Ramirez, Rev. Mex. Fis. \textbf{49}, 203 (2003).

\bibitem{Rahaman2010} F. Rahaman, K. K.  Nandi, A. Bhadra, M. Kalam, K. Chakraborty, Phys. Lett. B \textbf{694}, 10 (2010).

\bibitem{Potapov2016} A. A. Potapov, G. M. Garipova, K. N. Nandi, Phys. Lett. B \textbf{753}, 140 (2016).

\bibitem{Li2012} M. H. Li,  K. C. Yang, Phys. Rev. D \textbf{86}, 123015 (2012).

\bibitem{Xu2018} Z. Xu, X. Hou, J. Wang, Class. Quantum Grav. \textbf{35}, 115003 (2018).

\bibitem{Hou2018}  X. Hou. Z. Xu, J. Wang. J. Cosmol. Astropart. Phys. \textbf{12}, 040 (2018).

\bibitem{Haroon2019} S. Haroon, M. Jamil, K. Jusufi, K. Lin, R. B. Mann, Phys. Rev. D \textbf{99}, 044015 (2019).

\bibitem{Rizwan2019} M. Rizwan, M. Jamil, K. Jusufi, Phys. Rev. D \textbf{99}, 024050 (2019).

\bibitem{Hendi2020} S. H. Hendi, A. Nemati, K. Lin, M. Jamil, Eur. Phys. J. C \textbf{80}, 296 (2020). 

\bibitem{Zhang2021} H. X. Zhang, Y. Chen, T. C. Ma, P. Z. He, J. B. Deng, Chinese Phys. C \textbf{45}, 055103 (2021).

\bibitem{Atamurotov2021} F. Atamurotov, A. Abdujabbarov, W. -B. Han, Phys. Rev. D \textbf{104}, 084015 (2021).

\bibitem{Rayimbaev2021} J. Rajimbaev, S. Shaymatov, M. Jamil, Eur. Phys. J. C \textbf{81}, 699 (2021).


\bibitem{Atamurotov2022} F. Atamurotov, U. Papnoi, K. Jusufi, Class. Quantum Grav. \textbf{39}, 025014 (2022). 

\bibitem{Ndongmo2023} R. Ndongmo, S. Mahamat, C. B. Tabi, T. B. Bouetou, T. C. Kofane, Phys. Dark Univ. \textbf{42}, 101299 (2023).

\bibitem{Rakhimova2023} G. Rakhimova, F. Atamurotov, F. Javed, A. Abdujabbarov, G. Mustafa, Nucl. Phys. B \textbf{996}, 116363 (2023).

\bibitem{Qiao2023} C. -K. Qiao, M. Zhou, J. Cosmol. Astropart. Phys. \textbf{12}, 005 (2023).

\bibitem{Qi2023} S. Qi, Q. Li, Y, Zhang, Q. Q. Li, Mod. Phys. Lett. A \textbf{38}, 2350002 (2023). 

{
\bibitem{Heydarifard2023} M. Heydari-Fard, S. G. Honarvar, M. Heydari-Fard,  MNRAS \textbf{521}, 708 (2023).}

\bibitem{Das2024} A. Das, A. R. Chowdhury, S. Gangopadhyay, Class. Quantum Grav. \textbf{41}, 015018 (2024). 

\bibitem{Sadeghi2024} J. Sadeghi, S. N. Gashti, Phys. Lett. B \textbf{853}, 138651 (2024).

\bibitem{Ma2024} S. J. Ma, R. B. Wang, J. B. Deng, X. R. Hu, { Eur. Phys. J. C \textbf{84}, 595 (2024).}

\bibitem{Sood2024} A. Sood, Md S. Ali, J. K. Singh, S. G. Ghosh, { Chinese Phys. C  \textbf{48}, 065109 (2024).}



\bibitem{Biz2024} B. Hamil, B. C. L\"{u}tf\"{u}o\u{g}lu, Ann. Phys.   \textbf{472}, 169861 (2025). 

\bibitem{Heisenberg} W. Heisenberg, H. Euler,  Z. Phys. \textbf{98}, 714 (1936).

\bibitem{Schwinger} J. Schwinger,  Phys. Rev. \textbf{82}, 664 (1951).

\bibitem{Yajima} H. Yajima, T. Tamaki, Phys. Rev. D. \textbf{63}, 064007 (2001).

\bibitem{Kruglov} S. I. Kruglov, Mod. Phys. Lett. A. \textbf{32}, 1750092 (2017).

\bibitem{Ruffini} R. Ruffini, Y. B. Wu, S. S. Xue, Phys. Rev. D \textbf{88}, 085004 (2013).

\bibitem{Macias} N. Breton, C. Lammerzahl, A. Macias, Class. Quantum Grav. \textbf{36}, 235022 (2019).

\bibitem{Alfredo} M. Maceda, A. Macias, Phys. Lett. B \textbf{788}, 446 (2019).

\bibitem{Breton} D. Magos, N. Breton, Phys. Rev. D \textbf{102}, 084011 (2020).

\bibitem{Guerrero2020} M. Guerrero, D. Rubiera-Garcia, Phys. Rev. D \textbf{102}, 024005 (2020).

\bibitem{Amaro2020} D. Amaro, A. Mac\'ias, Phys. Rev. D \textbf{102}, 104054 (2020).

\bibitem{Allahyari2020} A. Allahyari, M. Khodadi, S. Vagnozzi, D. F. Mota, JCAP \textbf{02}, 003 (2020).



\bibitem{Breton2021} N. Breton, L. A. Lopez, Phys. Rev. D \textbf{104}, 024064 (2021).


\bibitem{Marco} M. Maceda, A. Macias, D. Martinez-Carbajal, Int. J. Mod. Phys. A \textbf{36}, 2150191 (2021).



\bibitem{Zeng} X. X. Zeng et al., Eur. Phys. J. C \textbf{82}, 764 (2022).

\bibitem{Sen} G. -R. Li, S. Guo,  E. -W. Liang, Phys. Rev. D \textbf{106}, 064011 (2022).

\bibitem{Qing} X. Ye, Z. Q. Chen, M. D. Li, S. W. Wei, Chinese Phys. C \textbf{46}, 115102 (2022).

\bibitem{Zhao} H. Dai, Z. Zhao, Sh. Zhang, Nuc. Phys. B  \textbf{991}, 116219 (2023).











\bibitem{Yu2023} Q. Yu, Q. Xu, J. Tao, Commun. Theor. Phys. \textbf{75}, 095402 (2023).

\bibitem{Rehman2023}  H. Rehman. G. Abbas. T. Zhu, G. Mustafa, Eur. Phys. J. C \textbf{83}, 856 (2023).

\bibitem{Magos2023} D. Magos, N. Bret\'on,  A. Mac\'ias, Phys. Rev. D \textbf{108} 064014 (2023).

\bibitem{Mushtaq2024} F. Mushtaq, X. Tiecheng, A. Ditta, A. Abduvokhidov, A. Asalkhon, New Astron. \textbf{108}, 102185 (2024).


  




\bibitem{Plebanski} J. F. Plebanski, Lectures on Nonlinear Electrodynamics. (Nordita, Copenhagen 1970).

\bibitem{Salazar} H. Salazar, A. Garcia D.,  J. Plebanski,  J. Math. Phys.  \textbf{28}, 2171 (1987).




\bibitem{Mehdi} S. G. Ghosh, S. D. Maharaj, D. Baboolal, T. H. Lee, Eur. Phys. J. C \textbf{78}, 90 (2018).

{
\bibitem{Ratra1988} B. Ratra, P. J. Peebles, Phys. Rev. D \textbf{37}, 3406 (1988).


\bibitem{Zeng:2020dco}
X. X. Zeng, H. Q. Zhang,  H. Zhang, Eur. Phys. J. C \textbf{80}, 872 (2020).

\bibitem{Zeng:2020vsj}
X. X. Zeng, H. Q. Zhang, Eur. Phys. J. C \textbf{80}, 1058 (2020).


\bibitem{Anacleto2021} M. A. Anacleto, F. A. Brito, J. A. V. Campos, E. Passos, Ann. Phys. \textbf{434}, 168662 (2021).

\bibitem{pal1}
R. Shaikh, K. Pal, K. Pal, T. Sarkar, MNRAS \textbf{506}, 1229 (2021).


\bibitem{Campos2022} J. A. V. Campos, M. A. Anacleto, F. A. Brito, E. Passos, Sci. Rep. \textbf{12}, 8516  (2022). 

\bibitem{Zeng:2021dlj}
X. X. Zeng, G. P. Li, K. J. He,
Nucl. Phys. B \textbf{974},  115639 (2022).


\bibitem{pal2} K. Pal, K. Pal, R. Shaikh, T. Sarkar JCAP \textbf{11}, 060 (2023).}

\bibitem{Anacleto2023} M. A. Anacleto, F. A. Brito, J. A. V. Campos, E. Passos, Eur. Phys. J. C \textbf{83}, 298 
(2023).

\bibitem{Ph24}
B. Hamil, B.C. L\"{u}tf\"{u}o\u{g}lu, Phys. Dark Universe  \textbf{44}, 101484 (2024).

\bibitem{Balendra} B. P. Singh,  S. G. Ghosh, Ann.  Phys. \textbf{395}, 127 (2018).

\bibitem{Carter} B. Carter, Phys. Rev. \textbf{174}, 1559 (1968).

{
\bibitem{Alex11} R. A. Konoplya, A. Zhidenko, Rev. Mod. Phys. \textbf{83}, 793 (2011).}


\bibitem{Schutz}  B. F. Schutz, C. M. Will, Astrophys. J. Lett. \textbf{291}, L33–L36 (1985). 

\bibitem{Mashhoon} B. Mashhoon,  1983 Proc. 3rd Marcel Grossmann Meeting on General Relativity ed H  Ning (Amsterdam: North-Holland) pp 599 – 608.

\bibitem{Blome} H. -J. Blome, B. Mashhoon, Phys. Lett. A \textbf{110}, 231 (1984).

\bibitem{Liu} H. Liu, B. Mashhoon, Class. Quantum Grav. \textbf{13}, 233  .(1996).

\bibitem{Iyer} S. Iyer, C. M. Will, Phys. Rev. D \textbf{35}, 3621 (1987).

\bibitem{Konoplya} R. A. Konoplya, Phys. Rev. D \textbf{68}, 024018 (2003).

\bibitem{Konoplyar} R. A. Konoplya, J. Phys. Stud. \textbf{8}, 93 (2004).

\bibitem{Matyjasek} J. Matyjasek, M. Opala, Phys. Rev. D \textbf{96}, 024011 (2017).

{
\bibitem{Konoplya13} R. Konoplya, A. Zhidenko,  A. Zinhailo, Class. Quant. Grav. \textbf{36}, 155002 (2019).}
 
\end{thebibliography}
\end{document}